\documentclass[preprint]{aastex}
\begin{document}

\title{Image-Subtraction Photometry of the Globular Cluster M3: 
identification of new double-mode
RR Lyrae stars}

\author{Gisella~Clementini}
\affil{INAF - Osservatorio Astronomico di Bologna, Via Ranzani 1, I-40127 
Bologna, Italy. 
email: gisella@sean.bo.astro.it} 

\author{T.~Michael~Corwin\altaffilmark{1}}
\affil{Department of Physics, University of North Carolina at Charlotte, 
Charlotte, NC 28223. 
email: mcorwin@uncc.edu} 

\author{Bruce~W.~Carney\altaffilmark{1}}
\affil{Department of Physics and Astronomy, University of North Carolina at 
Chapel Hill, 
Chapel Hill, NC 27599. email: bruce@astro.unc.edu}

\author{Andrew~N.~Sumerel}
\affil{Department of Physics, University of North Carolina at Charlotte, 
Charlotte, NC 28223. 
email: ansumere@uncc.edu}

\altaffiltext{1}{Visiting Astronomer, Kitt Peak National Observatory, 
National Optical Astronomy 
Observatories, which are operated by AURA, Inc., under cooperative agreement 
with the National 
Science Foundation.} 

\begin{abstract}
We have applied the image subtraction method (Alard 2000; Alard 
\& Lupton 1998) to the extensive M3 dataset 
previously analyzed by Corwin \& Carney (2001) using DAOPHOT and ALLSTAR. 
This new analysis has 
produced light curves and periods for fifteen variables not found in the 
previous study, but 
alread known to be variables (see Bakos et al.,
2000, catalogue), and
has also resulted in 
improved periods for several other variables. 
The additional variables recovered with the image subtraction analysis
are in the very central region of M3, where crowding 
is severe and the photometry was not of sufficient quality that 
it could be put on the standard system. 
The present study brings to 222 the total number of RR Lyrae 
variables in Corwin \& Carney (2001) 
M3 dataset, for which 
light curves and periods are available. Among them 
we have identified three new candidate double-mode pulsating variables 
(V13, V200, and V251) reported here 
for the first time. This brings to 8 the total number of double-mode RR Lyrae 
(RRd's) identified 
in M3. Of the newly discovered RRd's V13 is unusual in that it has the 
fundamental as the dominant pulsation mode.
M3 is unique among the 
globular clusters in having RRd variables with a dominant fundamental
mode.
Two of the new candidate RRd's (V13 and V200) have period ratios as low as 0.738-0.739.
They lie well separated from all previously known
double-mode variable stars in the Petersen diagram, in  positions
implying a large spread in mass and/or, less likely, in heavy element mass fraction, 
among the M3 horizontal branch (HB) stars. 
 We explore mass transfer and helium enhancement as possible explanations
for the apparent spread in HB masses. We also note that the masses derived 
from the double-mode analyses now favor little mass loss on the red giant
branch.

We find that 
V200 has changed its dominant pulsation mode from fundamental to first 
overtone, while V251 has changed its dominant mode from first overtone to
fundamental in the interval 1992 to 1993. Together with M3-V166
(Corwin et al.\ 1999) this is the first time that double-mode 
variables are observed to switch their dominant pulsation modes 
while remaining RRd's. The phenomenon is found to occur in a one
year time-span thus suggesting that these stars are undergoing a
rapid evolutionary phase, and that  
both redward and blueward evolution may take place among the 
horizontal branch stars in the Oosterhoff type I cluster M3.
 
The unusual behavior of the M3 RRd's 
is discussed in detail and compared to that of the 
double-mode RR Lyrae identified so far in globular clusters and in 
the field of our
and other Local Group galaxies.
 We find lack of correlation between the presence of RRd variables
and any of the cluster structural parameters.

{\it Key words\/}: globular cluster: individual (M3) 
--- RR Lyrae variables, double-mode pulsators --- stars: evolution 
--- stars: fundamental parameters --- stars: oscillations --- stars:
variables: other 
\end{abstract}

\section{INTRODUCTION}
M3 (NGC 5272) is among the most important and extensively studied Galactic globular
clusters. Often considered as a prototype of 
globular clusters of intermediate-poor metallicity, M3  
contains the largest number of RR Lyrae variable stars 
(N$_{\rm RR}$) within a single 
cluster (N$_{\rm RR} \geq$ 182, Clement et al.\ 2001) and is among 
the ten Galactic clusters 
with highest
specific frequency of such variables (${S_{RR}}$=49.0, from the 2003 
update to Harris, 1996, 
catalogue on Globular clusters, available at 
http://www.physics.mcmaster.ca/Globular.html). 

Since the pioneering study by Sandage (1953), 
M3 has been the subject of 
a very large number of photometric surveys. The most recent ones include 
Buonanno et al.\ (1994), Ferraro et al.\ (1997a), and studies based 
on Hubble Space Telescope (HST) data (Ferraro et al.\ 1997b,c, and 
Rood et al.\ 1999). 
The color-magnitude diagram 
(CMD) of M3 displays a horizontal branch (HB) spanning a very wide range in 
color and a quite narrow red giant branch (RGB;
Ferraro et al.\ 1997a).
The color of the RGB is 
known to depend on metal abundance (see for instance Renzini 1997, and Buonanno,
Corsi \& Fusi Pecci 1981).
Its
intrinsic width, after observational effects are removed, is
an indicator of metallicity spread and provides upper limits to the
dispersion of the elements having ``low'' ionization potential 
(e.g iron, see for instance Suntzeff 1993).
Thus the narrowness of 
the M3 RGB suggests a homogeneous iron abundance 
of the M3 stars.
Suntzeff (1993) reports an upper limit of ${\rm \sigma [Fe/H]} <$0.03 dex
to the metallicity dispersion in M3 based on a number of 
photometric and spectroscopic studies of the cluster.

Recent detailed spectroscopic abundance analyses based on high resolution
spectroscopy  
confirm the very low 
dispersion in iron abundance of the M3 stars, although some spread
exists among independent [Fe/H] estimates in the literature: 
[Fe/H]$_{\rm II}$= $-$1.50$\pm 0.03$ (Kraft \& Ivans 2003),
[Fe/H]= $-$1.49$\pm 0.02$ (from Kraft et al.\ 1999), 
$- 1.34 \pm$0.06 (Carretta 
\& Gratton 1997), $-$1.47$\pm 0.03$ (Kraft et al.\ 1992).
Spectroscopic observations also reveal
$\alpha$-capture 
elements enhancement by $\sim$+0.3 dex (Armosky et al.\ 1994, Carney 1996,
Salaris \& Cassisi 1996) and
star-to-star
variations and inhomogeneities in the abundances of the CNO group elements
(Suntzeff 1981, Norris \& 
Smith 1984, Kraft et al.\ 1992, Smith et al.\ 1996, Lee 1999,
Pilachowki \& Sneden 
2001) among the M3 giants, with both 
oxygen-rich ([O/Fe]$\simeq$=+0.3) and 
oxygen-poor ([O/Fe]$\simeq$=$-$0.15) stars coexisting in the
 cluster (Kraft et al.\ 1992). However, this is not in contrast 
with the small spread in color of the RGB
since these metals have ``high'' ionization potentials and their 
variation is not expected to spread out
significantly the RGB [Renzini 1977; Rood 1978 (unpublished)].

M3 contains an extremely rich population of variable stars
(N$_{\rm var}$=274 according to 
Bakos, Benko \& Jurcsik 2000) mainly 
consisting of RR Lyrae variables, but including also
SX Phoenicis stars, and a few long period variables 
(semi-regular and  W Vir stars).
The first modern studies of the variable star content of M3 date back to the
photographic surveys of
Roberts \& Sandage (1955), Baker \& Baker (1956), and Sandage (1959).
More recent studies include the CCD photometric surveys by Kaluzny et al.\ 
(1998) and Carretta et al.\ (1998; 60 variables), and the new catalogue by
Bakos et al.\ (2000)   
who presented improved identification and astrometry for all known or
suspected variables in M3. 
However, the most extensive study of the M3 variables is that of 
Corwin \& Carney (2001, hereafter CC01) who obtained $BV$ CCD photometry, 
light curves, and ephemerides for 207 of the RR Lyrae 
variables, and O$-$C diagrams for a subsample of 127 of them.
More recently, Strader, Everit 
\& Danford (2002), have 
presented an image subtraction analysis
(Alard 2000, Alard \& Lupton 1998) of the variables in the core of M3, 
adding 11 new candidates (among which 10 possible RR Lyrae stars)
to the already 
overwhelming list of variable stars in M3.

The total number of confirmed and/or suspected RR Lyrae stars 
identified in M3 by the above studies is larger than 230. Of them about
76 \% are fundamental
mode pulsators (RRab), 22 \% are first overtone pulsators (RRc), 
and 8 are double-mode variables (RRd), of which 3 were identified in
the present study. The transition between 
fundamental
and first overtone pulsators occurs at P$_{tr} \sim 0.45$ d
(see Figures 3 and 7 of CC01), and 
the average period of the fundamental mode pulsators is 
$<$P$_{ab}>$=0.561 d (CC01), thus making of M3 the best example of 
Oosterhoff (1939) type I (Oo I) clusters.  

In this paper we present a new analysis of the 
CC01 data using the 
image-subtraction technique 
(Alard 2000; Alard \& Lupton 1998).
The new analysis, described in Section 2,  
resulted in the recovering 
of additional variables not found by CC01, and in improved periods. 
The new period determination is 
described in Section
3. Three new candidate double-mode pulsating variables (V13, 
V200, and V251) were identified with the present study (Section 4). 
Their pulsational 
properties
are discussed in Section 5 where we also present a detailed comparison with 
other known cluster and field RRd's.

\section{IMAGE SUBRACTION PHOTOMETRY}

The photometric data used in the present study were obtained 
with the 0.9 m telescope at the
Kitt Peak National Observatory
in three 
observing runs spanning a period of five years (six nights in 1992, seven
nights in 1993, and one night in 1997). They consist of 190 $B$ and 189 $V$
frames with typical exposure times of 500 s and 300 s each, respectively.
A complete description of the observations and data processing 
can be found in CC01. 
In an attempt to improve the quality of the light curves and to recover 
known variables 
(see Bakos et al., 2000, catalogue) that had not been found by CC01, we employed the image subtraction 
package ISIS V2.1 (Alard \& Lupton 
1998; Alard 2000) on the extensive dataset of CC01. 

The ISIS analysis measures the difference in flux for stars in each 
image of the time series relative to their flux in a reference image 
obtained by stacking a 
suitable subset of images, and after convolving the images
with a kernel to account for seeing
variations and geometrical distortions of the individual frames.
We used the 10 $B$ and the 10 $V$ images with 
the best seeing of the 1992 and 1993 runs, respectively, 
to build up the reference images.
The new analysis has produced light 
curves and periods 
for fifteen 
variables 
(V180, V200, V210, V242, V244, V249, V250, V251, V253, 
V254, V255, 
V257, V264, V270, V271) that had not been found in CC01 study. 
The analysis has 
also resulted 
in improved periods for other variables. The recovered variables are all
located in the very central regions of M3.
In an attempt to convert the ISIS differential flux data to standard 
magnitudes we used 
DAOPHOT and ALLSTAR to obtain instrumental magnitudes for the new variables in 
the $B$ and $V$ reference images of the ISIS reductions. Due 
to crowding in the center, 
the photometry for the additional 15 variables not found by CC01 was 
unreliable. All had 
either high chi values, large magnitude errors or both. Almost all were 
overluminous 
for horizontal branch stars probably indicating that they are unresolved 
blends. No further 
attempt was made to put these stars on the standard system.

\section{NEW PERIOD DETERMINATIONS}

Periods for ten of the 15 variables we recovered were determined using 
the period-search program KIWI, kindly provided to us by Dr. Betty Blanco.
KIWI searches for periodicity by 
seeking to 
minimize the total length of the line segments that join adjacent observations 
in phase 
space, i.e., to maximize the smoothness of the light curve. 

Ephemerides for V250, V264, V270 and the three newly discovered double-mode 
RR Lyraes
(V13, V251, and V200) were derived with GRATIS (GRaphycal Analyzer of TIme 
Series) a private 
software developed at the Bologna  Observatory by P. Montegriffo, 
G. Clementini and L. 
Di Fabrizio (see Clementini et al.\ 2000).
GRATIS performs a period search 
according to two different algorithms :  (a) a Lomb periodogram  (Lomb 1976,
Scargle 1982) and (b) a best-fit of the  data with a truncated Fourier series
(Barning 1963). 
GRATIS also allows pre-whitening the data for the first periodicity, and 
searching for a second periodicity on the residuals with respect to the 
first periodicity best fit model. 

Periods and epochs of maximum light for the fifteen variables not present in CC01 are given 
in Table 1 along with
the new periodicities we derived for M3-V13, an RR Lyrae star reported to have
non-repeating light curves by CC01. We also provide in column 4 of the table the 
periods recently derived for some of these variables by Strader et al.\ (2002).
 
The data from our three observing runs cover 1867 days, spanning about 6900 
cycles for the 
shorter period variables, and about 2300 cycles for the longer period ones. 
Differential $B$ flux light curves for the new variables based on the 1993 
data and on the ephemerides  
given in column 2 and 3 of Table 1 are shown in Figure 1. 
The different symbols represent 
data from each of the seven nights of the 1993 run. 
The double-mode pulsators are plotted using the dominant 
period. Photometry on a calibrated magnitude scale is available only for V13.
In addition to the 15 variables recovered in this study that 
were not found by CC01,  
significantly improved light curves were obtained for many variables, 
some resulting 
in improved periods. Table 2 lists variables whose new 
periods differ by more than 0.0001 d 
from those given in CC01. The variable name in parentheses in 
Table 2 is the designation of 
that variable in CC01.

The CC01 analysis and our current study were either 
unable to obtain data for the following numbered variables listed in  
Bakos et al.\ (2000) or found them not to be variables.
The parenthetical comments refer to information from Bakos et al.\ (2000).  
V164 (possible variable of unknown type); V185 (variability not detected); 
V198 (variability not detected), CC01 did not observe this star 
and erroneously misclassified V245 as V198; 
V217 (RRab); V224 (variability suspected), this variable
is present in the CC01 and Strader et al.\ (2002) data, but both analyses 
failed to detect variability; V230 (unknown variable type); 
V233 (variability not detected); 
V237 (SX Phe); V238 (variability 
not detected); V262 (not distinguishable from V241 except by HST); 
V263 (SX Phe); V265 (RR Lyrae blended with V154 and V268); V266 (RRc); 
V267 (SX Phe); V268 (RR Lyrae blended with V154 and V265); V269 (RRc).

Strader et al.\ (2002) performed an image subtraction analysis 
of the variables in the core 
of M3. Their data consisted of 61 $V$ band images 
obtained over a five month interval in 2001. 
They provide a period for one RR Lyrae variable, V262, that was not found in either CC01 
or the current study. They also provide a period for the RRab V229 which was found 
in our current study, but whose light curve was extremely noisy. 
Their period of 0.6877 phases our data somewhat.
Strader et al.\ were able to find an additional nine variable stars not 
found in CC01 but that are part of the present analysis
(namely, V200, V242, V250, V253, V254, V257, V264, V270 and V271). They also 
provided improved periods for several variables listed 
in CC01. However, many of their 
periods did not phase our data properly. Strader et al.\ identified also 
eleven new 
suspected variables in M3, among which 10 were classified as RR Lyrae stars.
 Our data did not include their S1, S2, S5, 
S8 (a long period variable), and S10. Of the remaining suspected variables, 
there was too much scatter in our data to produce convincing light curves 
for any periods. There is some indication in our data that S3 
and S7 may be variable.

\section{DOUBLE-MODE PULSATORS}

Double-mode variables are stars that pulsate simultaneously in both the 
fundamental and the first overtone radial pulsation modes. First 
evidence for double-mode
pulsation was the discovery that the field RR Lyrae star AQ Leo 
(Jerzykiewicz \&
Wenzel 1977, Jerzykiewicz, Schult, \& Wenzel 1982) had a secondary 
first overtone 
periodicity beside its primary 
fundamental pulsation period. 
The first variable of this type in a globular cluster was discovered
by Goranskij (1981, V68 in M3). 
Since these early discoveries an increasing number
of double-mode RR Lyrae variables have been identified.
Sandage, Katem, \& Sandage (1981) found that some variables in the 
globular cluster M15 
exhibit unusually large scatter in their light curves and suggested they 
might be double-mode pulsators. Cox, Hodson, \& Clancy (1983), and 
Nemec (1985b) confirmed 
this finding and Nemec designated these stars RRd variables. 
Additional RRd variables have been found in a number of other globular 
clusters, among the 
field Milky Way (MW) RR Lyraes, and in several nearby galaxies. 

Double-mode RR Lyrae variables have great importance in constraining 
certain stellar 
parameters. It is well known that the pulsation periods depend on basic 
stellar parameters 
such as luminosity, effective temperature, mass, and chemical composition, 
most importantly 
on the metal content (Kovacs 2001b and references therein). 
RRd's can be used to estimate the mass and the 
mass-metallicity relation of HB stars. Masses of double-mode 
pulsators are evaluated from the ratio between the first overtone (P$_{1}$) and
the fundamental (P$_{0}$) pulsation periods through the so called 
``Petersen diagram''
(Petersen 1973). Pulsation models trace loci of constant mass in this diagram  
that plots the P$_{1}$/P$_{0}$ ratio versus P$_{0}$, from which stellar masses 
can be estimated.

The existence of double-mode pulsators suggests the possibility that these 
stars might be 
switching pulsation modes from overtone to fundamental or vice versa
while evolving across the HB instability strip. Thus they could provide 
information
on the direction and rate of evolution on the HB. Very recently Clementini
 et al.\
(2003b) found that the LMC  double-mode pulsators are systematically offset to 
slightly higher luminosities in the $V_{0}~~{\rm versus}~~$[Fe/H] relation 
defined by the
LMC RR Lyraes, thus providing support to the hypothesis that they may be more
evolved than the single-mode RR Lyrae stars.  
Cox et al.\ (1983) 
point out, however, that the theoretical switching time is only about 
150 years, much 
too short to account for the large observed number of RRd variables. 
Nonetheless, there is other evidence that the RRd 
behavior can change on
a short time scale (Clement et al.\ 1993, Jurcsik \& Barlai 1990).
The results presented in the present paper show that this is
in fact the case for many of the double-mode RR Lyrae in M3.
The physical mechanism for 
long-term maintenance of double-mode pulsation is still not well understood.

It is interesting to note that RRd variables occur only in a 
very small percentage 
of the Galactic globular clusters (GGCs) and appear also to be quite rare 
in the field of our Galaxy.
They are instead more frequent in the field of other Local Group dwarf galaxies,
where the new surveys based on large field telescopes and
the continuous monitoring of the microlensing studies are constantly  
increasing their number.

One RRd was identified in both NGC 2419 and NGC 6426 
 by Clement et al.\ (1993), 5 in M3 (Corwin, Carney \& Allen 1999 and reference
 therein) and 3 more
are reported in the present paper,
 12 in M68 (Walker 1994), 14 in M15 (Nemec 1985b), and 17 in IC 4499 (Walker 
and Nemec 1986).
Many other clusters have been searched with only negative results. 
Even the very RR Lyrae-rich cluster $\omega$ Cen has no RRd variables 
(Nemec et al.\ 1986). 
Only 6 RRd's were known so far in the field of our Milky Way 
(see Clementini et al.\
2000 and references therein), and the search in OGLE-II 
extensive database for variable stars in the Galactic bulge has added only 3 further
candidates (Mizerski 2003). 
However, Cseresnjes (2001) found 13 more Galactic RRd variables lying along the
line of sight to the Sagittarius dwarf galaxy. 

Turning to the extragalactic Local Group systems: 
6 RRd candidates are reported in Carina by Dall'Ora et al.\ (2003); 
10 RRd's have been identified in Draco (Nemec 1985a);
12 in the field of Fornax and 8 in the globular cluster Fornax 3 
(Clementini et al.\ 2003a) 
\footnote{The number of field RRd's in Fornax is likely 
a lower limit since 
these authors discuss only the variables they detected 
in 1/8 of the area of the galaxy they surveyed.}; 
18 in Sculptor 
(Kovacs 2001a); 40 in Sagittarius (Cseresnjes 2001); 
and 57 in the Small Magellanic Cloud (Soszy\'nski et al.\ 2002).
181 RRd's were found in the field of the Large Magellanic
Cloud by Alcock et al.\ (1997, 2000) on the basis of MACHO 
microlensing observations. This sample has been extended to
230 by Soszy\'nski et al.\ (2003) using OGLE-II database. The same 
authors also identified 6 and 2 RRd's respectively in the LMC globular clusters 
NGC1835 and 2019, while 4 RRd's are reported in Reticulum by   
Ripepi et al.\ (2003). 

In Tables 3, 4 and 5 we have collected information on the RRd variables 
identified so far in the various stellar systems.

Table 3 summarizes the literature data for the globular cluster RRd's.
In this table we provide: total number of RR Lyrae stars
in the cluster: N$_{\rm RR}$; number of double-mode RR Lyrae:
N$_{\rm RRd}$; cluster metallicity: [Fe/H] 
on the Zinn \& West (1984, hereafter ZW84) scale; 
classification of the clusters in Oosterhoff types (Oosterhoff 1939); 
specific frequency of the RR Lyrae variables: 
${\rm S_{RR}}$=
N$_{\rm RR} \times 10^{0.4(7.5+M_{\rm V})}$, 
which is the number of RR Lyrae stars in a cluster (N$_{\rm RR}$)
 normalized to 
a cluster absolute magnitude of $M_V=-7.5$;
horizontal-branch ratio: HBR=(B$-$R)/(B+V+R) where B and R are the 
numbers of HB stars respectively to the blue and to the red of the RR Lyrae
instability strip, and V is the number of RR Lyrae stars; 
and cluster structural parameters: 
central concentration: c=$log{\rm (r_t/r_c)}$; core radius: ${\rm r_c}$;
half-mass radius: ${\rm r_h}$, and tidal radius: ${\rm r_t}$) 
of the host clusters. Proper references for each of these quantities are 
given in the notes to Table 3.
Bragaglia et al.\ (2001, hereafter B01) have estimated
``pulsational'' masses for the Galactic cluster RRd's and for some 
of the field RRd's in the MW and in the LMC, 
using the Petersen diagram and an extension 
of Bono et al.\ (1996)  non-linear convective pulsational models.
Columns 12 and 13 of Table 3 provide the 
average mass of the MW cluster RRd's and its dispersion. These averages are the 
simple mean of the 
``pulsational'' masses 
derived for individual stars by B01.
Notice that masses being
derived for RR Lyrae stars from the new 
pulsation models are now
consistent with very little mass loss on the red giant branch.
That is, they are indistinguishable from those of
the main sequence turn-off.
Moreover, as already noticed by B01, the new models no longer predict a difference in 
mass between Oosterhoff type I and II clusters, or, depending on the 
adopted metallicity scale (ZW84 or Carretta \& Gratton 1997), the difference 
is opposite to what was previously thought (see B01 for details).

Table 4 summarizes information on the field RRd's identified so far in 
the various Local Group galaxies.
Metal abundances in Table 4 refer to the whole systems and not
to individual RRd's. They are on the ZW84 scale, and generally refer 
to the abundance of the
old stellar component in these galaxies, which in several cases
(e.g. Fornax, Sagittarius, etc.) is derived
from the pulsational properties of the fundamental mode RR Lyrae
stars.
For the LMC we list in Table 4 the
average metallicity recently derived from the spectroscopic
study of 100 LMC RR Lyrae stars
by Clementini et al.\ (2003b), and Gratton et al.\ (2003, in preparation). This value has been
transformed to the ZW84 metallicity scale according to Clementini et al.\ (2003b).
The Oosterhoff type classification  
of the Local Group galaxies has been assigned 
from the 
average period of the field ab type pulsators in each galaxy, according
to: Alcock et al.\ (1996) for the LMC, Soszynski et al.\ (2002) for the SMC,
Cseresnjes (2001) for
Sagittarius, Kaluzny et al.\ (1995) for Sculptor, Clementini et al.\
(2003a) 
for Fornax, Nemec (1985a) for Draco, and
Dall'Ora et al.\ (2003) 
for Carina.
For the Milky Way it refers to variables in the Galactic Center 
(Cseresnjes 2001).

Individual metal abundances are available only for 4 of the 
6 Milky Way field RRd's and
for 9 of the 230 
LMC RRd's, and these are provided in Table 5. 
These metallicities are either on the ZW84 scale, or on Clementini et al.\ 
(1995) scale. These two scales are quite similar to each other (B01).

At the moment there is no clear observational indication that the 
occurrence of the RRd phenomenon can be related to any specific cluster or 
stellar property (Kovacs 2001b). 
The RR Lyrae phenomenon is itself related to metallicity since RR Lyrae stars 
are rarely found in metal-rich clusters, and the frequency and mean periods
depend on metallicity among the metal-poor clusters.
The metallicity distribution of all known RRd's 
is shown in Figure 2.
In general only metal poor systems 
([Fe/H]$< -$1.5) 
seem to contain RRd variables, with only three of 
the field RRd's in the LMC with metallicity extending to values larger than 
this limit.  
The open triangle with the arrow in Figure 2
plots the total number of the LMC field RRd's assuming  
the average metal abundance of the LMC RR Lyrae in Clementini et al.\ (2003b),
transformed to the ZW84 scale.

As for the globular clusters, 
only a few clusters more metal rich than [Fe/H]=$-$1.0 contain RR Lyrae
stars. The 6 Galactic clusters containing RRd's  all 
are more metal-poor than 
[Fe/H]=$-$1.5, and 4 of them are at the very metal-poor edge of the 
Galactic globular cluster metallicity distribution ([Fe/H]$\leq -$2.0). 
Similarly, the 4 extragalactic globular clusters containing RRd's 
all are more metal-poor than 
[Fe/H]=$-$1.5.
This is  
shown in Figure 3, where we plot the 
specific frequency of the RR Lyrae variables: ${\rm S_{RR}}$, and 
the horizontal-branch ratio: HBR,   
as a function of the cluster 
metal abundance.  
Filled symbols mark the 
clusters where double-mode
RR Lyrae stars have been detected.
Some further parameter other than metallicity 
must determine why only these 10 metal-poor clusters
contain RRd's. The RR Lyrae specific frequency may be relevant, since 5 out 
of the
10 clusters hosting double-mode variables have ${\rm S_{RR}}$ values 
larger than 40, and two of them (Reticulum and IC4490) are among 
the clusters with highest RR Lyrae
specific frequency in general.
Seeking for correlations between number of double-mode variables and 
properties of the host cluster, 
in Figures 4 and 5 we plot the number of
double-mode variables (N$_{\rm RRd}$)
versus structural parameters of the 10 globular clusters 
containing RRd's. 
When transforming to parsecs the angular values of r$_c$, r$_h$, and r$_t$
we have assumed for the Galactic clusters the distances to the Sun
provided in Harris catalogue.
For the extragalactic clusters we have assumed distances of 51, and 138 Kpc for
the LMC, and Fornax dwarf spheroidal galaxy, respectively.
In the figures 
open and filled circles are used for the Milky Way Oosterhoff types I and II
clusters (Oosterhoff 1939), respectively; filled triangles represent 
the LMC
clusters, and the filled pentagon is Fornax 3. Among the extragalactic clusters 
containing RRd's 
Reticulum is 
of Oosterhoff type I  (Walker 1992), NGC 1835 is an Oosterhoff intermediate
(I/II, Soszynski et al.\ 2003),  Fornax 3 is an Oosterhoff type II cluster
(Clementini et al.\ 2003a),
while no classification is available 
yet for NGC 2019.
M15 is a core-collapsed
cluster and is labelled with a double c in the Figures.
No clear correlations are seen in either of these diagrams.


\subsection{NEW DOUBLE-MODE RR LYRAE STARS IN M3}
M3 was previously known to have 5 RRd variables: V68, V79, V87, V99, 
and V166 (Goranskij
1981, Cox et al.\ 1983, 
Nemec \& Clement 1989, 
Clement et al.\ 1997, and Corwin et al.\ 1999). Here we 
report the
discovery of 3 additional candidate RRd variables, V13, V200 and V251.

V13: This variable was discussed in CC01. As the article was going to press 
it was realized 
that V13 is probably a double-mode pulsator. A more detailed analysis 
using both the original 
data and the more recent image subtraction data has determined that 
the star has in fact two periodicities with a period ratio close to the 
canonical 
value for double-mode RR Lyrae stars, the period ratio range of all the 
presently 
known RRd stars being in the range
0.741 $<$ P$_1$/P$_0 <$ 0.748 (Popielski, Dziembowski, \& Cassisi 2000).

Figure 6 shows the periodograms for the DAOPHOT $V$ data of V13 
obtained with GRATIS using the Lomb algorithm to identify the most probable 
frequency of
the data on a wide interval of 0.22-0.8 d. Exactly the same periodicities 
are found 
in the full 1992 + 1993 +1997 data-set (top panel of Figure 6), and when the 
1992 (middle panel), and 
the 1993 + 1997 (lower panel) data subsets are analyzed separately.
These correspond to  
a dominant period of about 0.476 d (fundamental mode, frequency of 2.10 /d) 
and a secondary 
period of about 0.351 d (first overtone mode, frequency 2.85 /d). The 
frequency 3.10 /d is 
an alias of the fundamental mode and the frequency 1.85 /d is an alias of 
the first overtone 
mode. A more detailed determination of  
dominant and secondary frequency components was obtained by (i)  reducing 
the interval around the
primary periodicity and using GRATIS Fourier algorithm to find the best fit, 
and (ii) by
pre-whitening the data with the dominant primary frequency and searching for 
the secondary 
frequency component in the residuals with respect to the best fit model of 
the primary
frequency. The derived periods are ${\rm P}_0$=0.47951 $\pm$ 0.00001 d and 
${\rm P}_1$=0.35382 $\pm$ 0.00001 d
for a 
period ratio of 0.738. This value is lower than found so far for RRd 
variables in M3 and for RRd 
variables in general.

In Figure 7, the top panel shows the light curve of V13 using the 
DAOPHOT $V$ data with no 
pre-whitening and folded according to the highest amplitude, primary 
(fundamental) period of 
pulsation ${\rm P}_{0}$=0.47951 d. 
 The rms deviation from a 6 harmonics best 
fit model is 0.091 mag (much larger than expected from observational 
errors alone: 
0.02 mag in $V$), the amplitude of the light variation is 0.576 mag, 
and the intensity-averaged 
 $V$ magnitude is 15.649 mag. 
 The middle panel of Figure 7 shows the light curve of 
 the primary (fundamental) period of 
 V13 after pre-whitening of the secondary (first overtone) period 
${\rm P}_{1}$=0.35382 d.
 The amplitude of the light 
variation is A$_0$ = 0.534 mag and the average $V$ magnitude is 15.662 mag. 
The residuals 
from the 6 harmonic best fit model have been reduced to 0.046 mag.
 Finally, the bottom panel of 
Figure 7 shows the light curve of the secondary (first overtone) period after 
pre-whitening of the primary (fundamental) period.
The amplitude of the light curve of the secondary period is A$_1$ = 0.246 mag.
Figure 8 shows the light curves 
of V13 using the DAOPHOT $B$ data. The light curve  with 
no pre-whitening (top panel of 
Figure 8) has residuals from the 6 harmonics best fit model of 0.115 mag 
(to compare with 0.02 mag of the $B$ photometric errors) 
and residuals of 
0.051 after pre-whitening (middle panel). A$_0$ = 0.693 mag and A$_1$ = 0.308 
mag. 
The intensity-averaged $B$ magnitude, after pre-whitening is 
15.930 mag. 

A similar analysis of the Alard image subtraction data 
confirms the double-mode behavior of V13 and the periods given above for the 
dominant and secondary pulsations. 

As is usual in RRd 
stars the amplitude of the secondary mode is about half that of the dominant 
mode:
A$_1$/A$_0$=0.46 and 0.44 in $V$ and $B$ respectively.
However, contrary to the vast majority of the known RRd's,
but like M3-V68 in the time interval 1920-1926 (Nemec \& Clement 1989),
 in this star the primary
periodicity is the fundamental mode.
Indeed, only a very small percentage of the known RRd variables are found 
to have a dominant fundamental mode: AQ Leo in the Milky Way 
(Jerzykiewicz \&
Wenzel 1977, Jerzykiewicz, Schult, \& Wenzel 1982), 
and a few field RRd's in the LMC (Alcock et al.\ 2000). However, 
so far M3 is the only
cluster where this phenomenon has occurred.

V200: This variable was not found in the original CC01 analysis. The 
photometry for V200 
in the current analysis is not good enough to reliably determine standard 
magnitudes and 
the analysis of this variable was based on the Alard $B$ 
 differential 
flux data. This caused 
some uncertainty in the period search of the star.
 Figure 9 shows 
the periodograms for the 1992 + 1993 + 1997 data (top panel), the 1992 data (middle panel), 
and the 1993 + 1997 data (bottom panel)
obtained with GRATIS 
searching the data with the Lomb algorithm on a wide interval of 0.22-0.8 d. 
The top panel of Figure 9 shows  that there are two frequencies of about 
equal strength, one of about 0.488 d (fundamental mode, frequency of 2.05 /d) and another 
of about 0.357 d (first overtone mode, frequency 2.80 /d). The frequencies 3.05 and 4.05 /d 
are aliases of the fundamental period and the frequencies 1.80 and 3.80 /d are aliases of 
the first overtone period. A 
further search using the Fourier algorithm and pre-whitening 
the data with the 
dominant primary periodicity
 gives most likely periods of 0.4867 d 
and 0.3598 d, for a period ratio of 0.739. This value is lower than expected for RRd 
variables in M3 and for RRd variables in general, but slightly higher than V13 discussed above.
It should be noted, however, that period search solutions providing fundamental and first
overtone periodicities with period ratios as short as 0.735-0.734 d are also
compatible with the differential flux data of V200. However, the first solution is 
preferred since it provides a period ratio closer to the canonical values
observed for RRd stars.

A most important feature in Figure 9 is that 
when the 1992 and 1993+1997 data are analyzed separately it is found that in 1992 the fundamental 
and first overtone frequencies are of about equal strength
(see middle panel of Figure 9). However, in the 1993 + 
1997 data, the 
first overtone frequency is clearly dominant (see bottom panel of Figure 9).
This seems to suggest that V200 may have switched its dominant pulsation mode from
fundamental to first overtone.
A very similar behavior was observed in M3-V166 by 
Corwin et al.\ (1999) who employed a different period search algorithm 
than here, to study the
star, the phase dispersion minimization technique (PDM) developed by
Stellingwerf (1978).

The top panel of  Figure 10 shows the light curve of V200 
using the full 1992 + 1993 + 1997 Alard differential flux data 
with no 
pre-whitening and folded according to the highest amplitude, 
primary (first overtone) period of 
pulsation ${\rm P}_{1}$=0.3598 d. 
The middle panel shows the light curve of 
 the primary (first overtone) period of  V200 after pre-whitening of the secondary 
 (fundamental) period ${\rm P}_{0}$=0.4867 d. 
 Finally, the bottom panel 
shows the light curve of the secondary (fundamental) period after 
pre-whitening of the primary (first overtone) period.

V251: As was the case for V200, this variable was not found in the original CC01 analysis and the 
photometry in 
the current analysis is not good enough to reliably determine standard magnitudes 
for V251. Thus, also for this star the period search
was based on the Alard $B$ differential flux data.  
However, contrary to V200, for V251 it was possible to derive definite 
periodicities.
Figure 11 shows the 
periodograms for the 1992 + 1993 + 1997 data (top panel), the 1992 data (middle panel), and 
the 1993+1997 data (bottom panel) of the star obtained by GRATIS using the Lomb algorithm.
The top panel shows a dominant period of about 0.474 d 
(fundamental mode, frequency of 2.11 /d) and a secondary period of about 0.353 d 
(first overtone mode, frequency 2.81 /d). The frequency 3.11 /d is an alias of the 
fundamental mode and the frequencies 3.81 and 1.81 /d are aliases of the first overtone mode.

The dominant periodicity, best fitting the data with a 5 harmonics Fourier series,  
is ${\rm P}_{0}$=0.47423 $\pm$ 0.00001 d. Data pre-whitened according to the above primary
period are then best fitted by a two harmonics Fourier series corresponding to a 
secondary periodicity of ${\rm P}_{1}$=0.35384 $\pm$ 0.00001 d.
The corresponding period 
ratio of 0.7461 is in the range typical of RRd variables in M3. 

When the 1992 and 1993+1997 data are analyzed separately it is found that in 1992 the dominant 
mode was the first overtone for ${\rm P}_{1}$=0.35384 d, while in 1993+1997 
the dominant mode is the fundamental one
${\rm P}_{0}$=0.47423 d 
(see Figure 11 middle and bottom panels). Like V13, the present primary 
periodicity of V251 is the fundamental mode, however the behavior of the 
periodograms suggests that V251 may have switched its dominant pulsation mode from
first overtone to fundamental in the interval 1992 to 1993.

The top panel of  Figure 12 shows the light curve of V251 
using the full 1992 + 1993 + 1997 Alard differential flux data 
with no 
pre-whitening,  and folded according to the highest amplitude, 
primary (fundamental) period of 
pulsation ${\rm P}_{0}$=0.47423 d. 
The middle panel shows the light curve of 
 the primary (fundamental) period of  V251 after pre-whitening of the secondary 
 (first overtone) period ${\rm P}_{1}$=0.35384 d, and 
finally, the bottom panel shows the light curve of the secondary 
(first overtone) period after 
pre-whitening of the primary (fundamental) period.

Fundamental periods (P$_0$) and period ratios (${\rm P}_{1}/{\rm P}_{0}$) for
the 8 RRd's in M3 are summarized in Table 6 along with intensity-averaged
magnitudes and colors (columns 5 and 6) taken  from CC01, and 
distances from the cluster center (column 7).
 In column 4 of the Table we also list pulsational masses derived by 
B01 for the already known M3 RRd's, and rough estimates of the mass of the 
newly discovered RRd's, derived as described in Section 5.1.

\section{DISCUSSION}

M3 seems to be somewhat unusual with respect to its double-mode pulsators. 
We summarize here the distinctive characteristics of the cluster's RRd
variables.
The cluster is unique 
among the globular clusters in that 
it has RRd variables with a dominant fundamental mode: V13, V166
 (in 1992, but not 1993), V251 (in 1993, but not 1992), and possibly V68 
(see discussion in Corwin et al.\ 1999).
Of them V251 is located in the cluster core, while the remaining three
do not have a preferred physical location in the cluster, or occupy 
any different part of the CMD. 
Only 
very few RRd variables are presently found to have a
dominant fundamental 
mode, and they all are field variables 
(AQ Leo, and a few RRd's in the LMC, Alcock et al.\ 2000). 

M3 is so far the only cluster in 
 which RRd variables have been observed to switch from one dominant 
mode to 
another 
 while remaining RRd variables (Corwin et al.\ 1999 and the present 
analysis). 
Variations in the double-mode behavior have been observed in 
a number of RRd variables in M68 (V33 and V21, Clement et al.\ 1993), 
M15 (V30 and V53, Purdue et al.\ 1995, V26, Jurcsik \& Barlai 1990), 
M3 itself (V79, Clement et al.\ 1997),  
and in the Galactic field RRd AQ Leo 
(Jerzykiewicz, Schult, \& Wenzel 1982).
Most of these variations concern (i) initiation or cessation of the
double-mode behavior, and (ii) changes in 
the length and amplitude ratios of the two periodicities. 
However, none of these studies reports 
a switch between dominant modes to have actually
occurred in these stars. We will come back 
to this point in Section 5.2. 

M3 also has an unusually large range of period ratios, not only for a 
single cluster but 
for RRd variables in general.
It was generally assumed that RR Lyrae stars which pulsate simultaneously 
in the fundamental and first overtone modes have period ratios 
P$_1$/P$_0$ between 0.73 and 0.76 d (Cox et al.\ 1983, Nemec 1985b).
 Kovacs (2001a,b) reports that all known RRd stars 
have period ratios 
in the range of 0.741 and 0.748, and that variables within a single 
cluster cover only 
narrow ranges within this interval. In general, metal-poor clusters have 
RRd stars with 
larger period ratios and periods, while for metal-rich clusters these 
values are smaller. 
M3 does not seem to conform to this rule. Six of the eight RRd variables 
have period ratios 
in the range 0.743 and 0.747 (Corwin et al.\ 1999 and the current analysis) 
while the other 
two, V13 and V200 have period ratios of 0.738 and 0.739 respectively. 

An important question is whether, with proper observations, anomalies similar
to those seen in some of the M3 RRd's could be found also in other 
cluster and field RRd's. In this respect we note that assets of 
the present analysis were 
 the relatively extensive and widespread 
time coverage (6, 7 and 1 consecutive nights, arranged in a 6 years time span), 
the rather large number (190 $B$ and 189 $V$) and good photometric precision 
(0.02 mag in both $B$ and $V$ passbands)  
of our CCD data for the M3 variables, and the use of a powerful detection technique 
(e.g. the image subtraction method, Alard 2000, Alard \& Lupton 1998).
Thus it may be possible that with further observations 
and the use of appropriate detection techniques 
allowing to search the crowded cluster cores, new RRd variables and perhaps 
 some of the 
``anomalies" found among the M3 RRd's could also be detected 
in other GCs. 
For instance, it is interesting that the very extensive photometric survey of 
variable stars resulting from the MACHO project (Alcock et al.\ 1996) succeeded in
identifying 
some RRd's with dominant fundamental pulsation mode (Alcock et al.\ 2000,
and references therein) in the LMC.

\subsection{THE PETERSEN DIAGRAM}

Figure 13 shows the position of the M3 RRd's in 
the Petersen diagram of all known double-mode RR Lyrae stars identified 
so far in 
various different
stellar systems, for which periods and period ratios are found in the
literature.
Filled circles are used for the 3 new candidate 
double-mode stars in M3 
while filled squares mark the 5 previously known RRd's.
 Lines show the pulsational models
by Bono et al.\ (1996), which correspond to Z=0.0001, and M/M$_{\odot}$=0.65, 0.75, 
and 0.80, and their extension to larger mass and metallicity ranges
(M/M$_{\odot}$=0.85, and Z=0.0002, 0.0004, 0.0006, 0.0008) obtained by B01. The derivation of these
new models is fully discussed in B01, to which the interested 
reader is referred for further details.
Figure 13 provides a summary of our present knowledge of
both  ``observational'' (periods and period ratios) and ``theoretical''
(pulsational models) Petersen diagram.

Data plotted in Figure 13, which is an updated version of B01 Figures 5 and 6, 
 were taken 
from Garcia-Melendo \& Clement (1997) for NSV09295; 
from Clement et al.\ (1991, 1993) for AQ Leo, VIII-10, and VIII-58; 
from Clementini et al.\ (2000) for CU Com;
from Moskalik \& Poretti (2003) for BW7 V30; 
from Cseresnjes (2001) for the field MW RRd's in the
direction of Sagittarius  (13 RRd's);   
from Clement et al.\ (1993) for NGC 2419 (1 object), and NGC 6426 (1 object); 
from Corwin, Carney \& Allen (1999) for M 3 (5 objects); 
from Walker \& Nemec (1996) for IC 4499 (17 RRd's);
from Walker (1994) for M 68 (12 RRd's); 
from Nemec (1985b) for M 15 (14 RRd's); 
from Nemec (1985a) for Draco (10 RRd's); 
from Kovacs (2001a) for Sculptor (18 objects);
from Cseresnjes (2001) for Sagittarius (40 RRd's);
from Soszy\'nski et al . (2002) for the SMC (57 RRd's); 
from Soszy\'nski et al . (2003) for the LMC field RRd's (230 objects) and 
for the RRd's in the LMC clusters NGC 1835 (6 objects) and NGC 2019 (2 RRd's);
from Ripepi et al.\ (2003) for the 
Reticulum cluster (4 RRd's).
The 6 RRd candidates in Carina, and  
the field and cluster RRd's in Fornax (12 and 8 objects,
respectively), 
are not  shown in the figure since
periods are not available yet for these stars.

The large spread in period ratio of the M3 double-mode pulsators, a spread 
much larger than found in any other of the stellar systems containing 
RRd's, can clearly be seen in the Figure. 
Figure 14 shows an enlargement of Figure 13 in the region relevant for the M3
RRd's. Only the M3 double-mode variables are displayed now,
and lines of pulsational models corresponding to various values of the heavy element 
mass fraction Z are labelled
by mass and luminosity level, in solar units. 
Since both mass and metal abundance affect the predicted Petersen diagram, 
the dispersion of the M3 double-mode variables  
in Figure 14 can be explained either by a 
spread in heavy element 
mass fraction Z, or by a spread in mass.

B01 made the following assumption: (i) [Fe/H]=$-$1.66 (on the ZW84 scale) 
and $Z_{\odot}$=0.019, (ii) no significant spread in [Fe/H], as reasonable 
for stars within
a cluster and as confirmed by both spectroscopic determinations\footnote{ 
Spectroscopic abundance analyses find a small dispersion
in the [Fe/H] abundance of M3, independently of 
the adopted value and metallicity scale. For instance, Ivans \& Kraft (2003) 
find $\sigma$[Fe/H]$\le$0.03 dex, based on the analysis of 17 M3 red 
giants.}
 and 
the very low intrinsic width of the M3 Red Giant Branch
(see Introduction),  
and (iii) no $\alpha$-enhancement. They derived masses with 
dispersion of about 0.06 ${\rm M_{\odot}}$ around the average value 
M=0.779 ${\rm M_{\odot}}$, for the 5 previously known RRd's in M3. 
This dispersion is 3 times larger than they 
find for the double-mode variables in IC4499 and M68, and 1.5 larger than in M15 (see
column 13 of Table 3).

With a period ratio of P$_1$/P$_0$=0.7461 and a fundamental mode period 
$P_0$=0.47423, in the Petersen diagram  V251 falls 
very close to 4 of the  M3 
RRd's analyzed by B01, 
namely: V99 (P$_1$/P$_0$=0.7468, M/${\rm M_{\odot}}$=0.846); V166 
(P$_1$/P$_0$=0.7459, M/${\rm M_{\odot}}$=0.803);
V87 (P$_1$/P$_0$=0.7458, M/${\rm M_{\odot}}$=0.795); and V79 (P$_1$/P$_0$=0.7453, 
M/${\rm M_{\odot}}$=0.772). Thus according to B01
analysis V251 could have a mass around $\sim$0.8 ${\rm M_{\odot}}$.

With their 
significantly lower period ratios,
V13 (P$_1$/P$_0$=0.738) and V200 (P$_1$/P$_0$=0.739) lie well separated
from all the known 
RRd's and in-between the Z=0.0008, M/M$_{\odot}$=0.65, 
$\log L/L_{\odot}$=1.61, 1.72  model lines. Thus the comparison with the 
nonlinear 
pulsational models seems to suggest that V200 could be a $\sim$ 0.65 
M$_{\odot}$ 
pulsator
 with a luminosity level
intermediate between 1.61 and 1.72 and a total heavy-element mass fraction 
Z$\sim$0.0008.
V13 could be
either a lower mass variable (M $\sim 0.55$ M$_{\odot}$) with the same metal
abundance of V200, or a more metallic pulsator (Z$\sim$0.001) with mass 
similar to V200\footnote{If the above mass evaluations are correct, the average
mass spread of the M3 RRd's would be as large as $\sim$0.10 M$_{\odot}$.}. 
The higher
global abundances of V200 and V13 should be the result of a non
negligible and varying $\alpha$-enhancement ($\alpha$=0.2$-$0.4 dex for 
Z$_{\odot}$=0.02)
in these two variable giant stars.
A varying $\alpha$-enhancement and variations in oxygen abundance
in particular would not be in  contrast to the absence of star-by-star scatter 
in [Fe/H], since,  
as discussed by VandenBerg (1992) and Rood \& Croker (1985), variations
in oxygen do not affect the radii, hence the colors of the low-mass, 
low-metallicity
red giants. Such variations are expected, however, to affect the effective 
temperature hence color of stars near the main-sequence turnoff, though not to
an easily observable extent. However, observations show that the M3 giants 
have an almost constant 
$\alpha$-enhancement
(O, Si, Ca, Ti) by about 
0.3 dex (Armoski et al.\ 1984, Carney 
1996, Salaris and
 Cassisi 1996, and references therein). 

One may wonder whether  
the star-to-star 
variations in the elements of the CNO group 
reported in M3 by  
several independent spectroscopic studies 
may produce a heavy element mass fraction variation among the M3 horizontal 
branch (HB) stars.
%
%
Variations in both C and N abundances and carbon-nitrogen anticorrelation,
with a constant total abundance of C+N, 
were first discovered among the M3 giants by Suntzeff (1981), and later
confirmed by Norris \& Smith (1984) and Lee (1999). 
Oxygen variations with both oxygen-rich [O/Fe]$\simeq$=+0.3 and 
oxygen-poor [O/Fe]$\simeq$=$-$0.15 stars coexisting in the cluster, 
and oxygen-sodium anticorrelation 
(with mainly O rich and Na poor stars among the M3 giants), 
were found by Kraft et al.\ (1992). 
Finally, Smith et al.\ (1996) 
find anticorrelation between CN band strength and both 
[O/Fe] and [C/Fe] abundances, coupled with CN-[N/Fe] correlation, but they 
also show that the
total [(C+N+O)/Fe] is constant among the M3 giants.

 Thus the 
chemical 
mix is changing,  
 but the total heavy element mass fraction, Z, is probably
not changing, making very unlikely 
to explain the position of V200 and V13 in the Petersen diagram
in terms of a difference in Z.
Clearly, high resolution spectroscopy and direct measure of the 
elemental abundances  
 of the M3 double-mode pulsators are highly desirable in order to 
break the 
mass-metallicity
degeneration in the Petersen diagram, due to the similar effects of the 
metallicity 
and mass on the period ratio.



Similarly, if some
sort of mass transfer is likely to cause these low masses, 
then radial velocity monitoring should be able 
to reveal possible binary systems. 
Gunn \& Griffin (1979) and Pryor, Latham \& Hazen (1988) have searched 
for binaries among the red giants in 
M3.
Neither study found a high binary fraction. However, a re-analysis
of the data from Pryor et al.\ is in progress.
It may be
daunting to reveal binarity in the presence of pulsation, especially
multimode pulsation, but periodograms ought to pull out
a period well enough, and if photometry is used to monitor
the changes in temperature and radius, one might still be
able to obtain an orbital solution. The orbital period is likely to be
much longer than the pulsational periods and hence be extractable via Fourier
analysis technique\footnote{The RR Lyrae stars are descendents of red
giants whose radii may have been as large as 100 R$_{\odot}$, and stellar 
companions capable of accepting mass from the red giant (pre-RR Lyrae star) 
would likely have had orbital periods of weeks to months to, perhaps,
years.}.

Also further photometric observations of 
the M3 RRd's with improved spatial
resolution and covering larger time intervals are needed to remove the 
uncertainties still present in the period determinations (particularly 
for M3-V200)
and to monitor period/mode changes in these stars.

On the other hand, the very wide range in color (temperature) of M3
horizontal branch shows that 
some spread in mass exists 
among the M3 HB stars.
CC01 find also that the HB of M3 has a large vertical height of the
order of about 0.3 mag with a number of variables both subluminous and
overluminous than the zero age horizontal branch (ZAHB), that they
set at $V$=15.72 mag. 
In fact, 12 RR Lyrae stars in CC01 sample lie 
 below the ZAHB at an intensity-averaged $V$
 magnitude around $V \sim$15.826 mag, and 30 variables lie 
 above the ZAHB at an intensity-averaged $V$
 magnitude brighter than $V \sim$15.55 mag.
CC01 interpret these differences as possibly produced by both 
mass dispersion and evolutionary effects occurring among the M3 RR Lyrae stars.

 The subluminous RR Lyrae stars are located within
 81 arcsec from the cluster center, 
thus CC01 discuss whether a lower HB core mass resulting from
 star-star interactions in the high density cluster central region might
 be responsible for the lower luminosity of these stars.
Our results on the M3 double-mode pulsators, 
with two of the 
newly discovered RRd's being significantly less massive than 
the other M3 RRd's, seem to confirm that an unusually large spread in mass
is likely to exist within 
the M3 HB stars, and in particular in the very narrow region of 
the HB instability strip corresponding to 
the transition between fundamental and first
overtone pulsators.
M3 has also many blue stragglers, 
especially toward the center (Ferraro et al.\ 1997c)
where binaries are likely to be more common. 
Thus one possibility is whether the subluminous RR Lyrae and the 
small mass RRd variables could have acquired atypical masses from some
phenomenon related to the core properties of the cluster.
We have no radial velocity information but there are suggestions
that mass-transfer binaries may be more common in clusters' cores 
(Hut et al.\ 1992).

Among the two new candidate RRd's with possibly unusual small masses 
only V13 has calibrated photometry so we can check 
its position on the CMD.
According to the intensity-averaged magnitudes listed in Table 6
the RRd variables with calibrated photometry have average luminosity
$<V_{RRd}>$=15.63$\pm$0.04 and
lie close and slightly above the ZAHB level ($\delta V_{RRd}$=
$<V_{RRd}> - <V_{ZAHB}>$=$-0.09 \pm 0.04$ mag. In particular,  
V13 ($<V_{int}>$=15.612 and 15.662 mag in CC01 and in the present
analysis, respectively) lies about 0.1 mag above the ZAHB. Thus 
if the  star is actually less massive,  
evolution (or some other mechanism/parameter) must have compensated 
in part the lower luminosity 
corresponding to its lower mass. We will come back to this point at the 
end of this section.

\subsubsection{MASS-TRANSFER IN BINARY SYSTEMS}
If according to the Petersen diagram in Figures 13 and 14, the atypical RRd's
in M3 are those with unusually small period ratios, that  
 fall in regions where no other RRd has ever been found before
(namely V200, V13 and to a lesser extent V68), 
then according to the pulsational models these stars should have 
``smaller''
masses. We are assuming here that the 
Petersen diagram is applicable and that it is actually able to predict correct masses 
for stars undergoing the rapid evolution causing the period changes and the
mode switching 
observed in the M3 RRd's.  
A further caution is that we may also need improved periods and
period ratios for some of our stars, like V200 and V251.

In any case, if mass-transfer in a binary system 
triggered by the high central density of M3 core 
is the mechanism causing the spread in mass,  
these atypical RRd's 
should be the 
``donors''
of mass of such systems.
However, not all of the unusual M3 stars are in the core. 
Neither are all the blue stragglers, either. 
In fact, of the 3 RRd's with atypically short period ratios,
hence atypically small masses (V13, V200, and V68), only V200 
is in the central regions of the cluster (see column 7 of Table 6).
V251 is in the core region but it has 
a period
ratio, hence mass, similar to the other ``typical''
RRd's in M3, while V13, ``atypical'' period ratio and mass it is not in
 the cluster core.
 

Of the GGCs containing double-mode variables M68 also has a widely spread 
HB with a well extended blue tail and contains several likely blue 
stragglers (Walker 1994).
However, its RRd stars show a very modest mass dispersion
(${\rm \sigma_{M}} = 0.019$M$_{\odot}$); see column 13 of Table 3). 
Conversely, double-mode variables in M15,  
a core-collapsed cluster with only a few red HB stars, many variables, 
and a bifurcated blue HB 
with a long tail of very blue stars, exhibit mass dispersion 
(${\rm \sigma_{M}} = 0.036$M$_{\odot}$) intermediate
between M68 and M3.
Why do we see a spread in mass in M15 and M3 (clusters with very
different HB morphologies) and do not see it in M68 whose HB morphology is
intermediate between M3 and M15?

\subsubsection{HELIUM ENHANCEMENT}
An alternative possible explanation for the spread in mass and luminosity 
of the M3 RR Lyrae stars could be the presence of a spread in helium
abundance within the framework recently suggested by D'Antona et al.\ (2002).
These authors present stellar evolution models for globular cluster stars with
helium enhancement arising from self-pollution of the intracluster material 
and of the pre-existing main sequence structures by the ejecta of first 
generations of massive asymptotic giant branch (AGB) stars. As first suggested
by Norris et al.\ (1981) the self-pollution mechanism due to intermediate mass
stars would produce the CNO abundance spread and chemical peculiarities observed
in cluster stars, but as recognized by D'Antona et al.\ it 
would also cause a {\it helium enrichment}. 
Isochrones for GC ages computed by D'Antona et al.\ (2002) show that the most
compelling effect of the helium enhancement would be to lower the masses (by up to
about 0.08 M$_{\odot}$ for a helium enhancement Y from 0.23-0.24 up to 0.28) 
of the evolving red giant branch (RGB) stars at the RGB tip, as a result of 
both evolution and mass loss occuring in the RGB phase.
Thus the effect of a distribution in the initial value of the helium abundance
would best be seen on the HB where helium-enhanced HB and RR Lyrae stars   
would, on average, be more luminous and bluer, with a substantial thickening 
of the
HB, than stars with standard low helium abundance.
The helium enhancement should not affect, however, the periods of the RR Lyrae
stars, since, as shown by Caputo, Santolamazza \& Marconi (1998), for fixed
values of the structural parameters (mass, luminosity and effective temperatures)
the periods of fundamental and first overtone pulsators do not vary for a variation
of the helium content in the range 0.24$\leq$Y$\leq$0.31 (see Figure 2b of
Caputo et al.\ paper).

 The Y enhancement would also 
produce variations of the luminosity and color of the Main Sequence (MS), 
Turn-Off (TO) and RGB.
D'Antona et al.\ (2002) estimate that at Z=0.0002 
and for a helium enrichment from 0.24 to 0.28 the TO magnitude of the helium 
enriched stars would be 0.07 fainter, the
main sequence location would be 0.13 mag fainter, and both the color 
along the RGB and below the MS TO would be 0.015 mag bluer. We note that the 
spread in luminosity and color observed in the subgiant region of M3 (see Figure 4 of CC01, 
and Figure 15 of Ferraro et al.\ 1997a) could thus be a signature of helium
dispersion among the M3 stars, although, as discussed by Ferraro et al.\ 
(1997a), it could also
be the result of {\it optical blending} between MS-stars and between 
subgiant branch objects and blue stragglers.

Depending on the actual distribution of the helium enriched stars and on the 
total
amount of assumed average mass loss, the mechanism proposed by 
D'Antona et al.\ (2002) would explain the
various different HB morphologies, including the HB gaps and tails observed in
GCs, for values of the parameters largely in the observed range.
Of particular interest is the simulation of M3-like HB morphologies
(see top right panel of Fig. 4 in D'Antona et al.\ paper), where both 
helium-enriched and low-helium stars enter the RR Lyrae strip and well reproduce a
large fraction of the vertical height observed in the M3 HB. We specifically 
note that 
the predicted total mass spread of 0.08 M$_{\odot}$ (for a helium enrichment 
up to 0.04) would be consistent with  
both the spread in mass ($\sigma_{\rm M}$=0.06$-$0.10 ${\rm M_{\odot}}$) and the 
average overluminosity 
($\delta V_{RRd}$=$-$0.09 mag) observed in the M3 RRd variables.

On the other hand, since in D'Antona et al.\ (2002) scenario the helium 
enrichment is
induced by an intracluster mechanism (self-pollution), whose amount and 
extent may 
differ from cluster to cluster depending on ``local" conditions as for
instance the cluster metal abundance, total mass and concentration, 
with a proper fine tuning of the involved parameters it should be conceivable 
to explain, within the same framework, both the different HB morphology and 
the smaller spread in mass of the M15 RRd's, and the lack of mass spread 
observed among the M68 double-mode stars. 

\subsection{PERIOD CHANGES AND MODE SWITCHING}
\subsubsection{INTRODUCTION}
Changes in the pulsation period are a very common phenomenon among the 
RR Lyrae stars, and there is growing observational evidence that 
they occur also among the double-mode RR Lyrae variables.
Our results on the new double-mode RR Lyrae stars in M3 confirm this occurrence. 

Since RR Lyrae star periods are sensitive to the RR Lyrae star 
densities, hence luminosities, by the Ritter relation $P\sqrt\langle\rho\rangle =
Q$, and van Albada \& Baker (1971) basic equations of stellar
pulsation $P = f(M,L,T_{\rm eff})$, changes in the period mirror 
changes in the structure of the star, 
and thus are related to evolutionary effects. 
The period changes of the RR Lyrae stars thus 
represent a powerful tool for studying 
the evolution of the HB stars.
In particular, period increases should indicate redward evolution, and 
decreases 
should indicate blueward evolution, while the rates of change can provide 
information about the timescales of the evolutionary phenomenon. 

Evolutionary models  
(Sweigart \& Renzini 1979; Lee, Demarque, \& Zinn 1990)
predict that the RR Lyrae variables in the Oo I
clusters 
enter the instability strip as fundamental mode pulsators during their ZAHB
phase and evolve from red to blue, therefore are expected to have 
decreasing periods. RR Lyrae variables in the Oo II
clusters 
enter the strip as first overtone pulsators and evolve from blue to red, 
therefore are expected to have increasing periods.
However, the theoretical evolutionary HB models do not adequately reproduce 
either the
direction (increasing or decreasing period) or the rate of period changes
observed in RR Lyrae stars, since both positive and negative changes are
observed within RR Lyrae stars in a given cluster and, in absolute values,
the observed rates of changes are sometime one or two orders of magnitude
larger than predicted by evolutionary models.
Iben \& Rood (1970) concluded that both
the size and the patterns of the period changes exclude a simple
evolutionary explanation, as have others since (e.g. Clement et al.\ 1997, Alcock
et al.\ 2000). 
Indeed, magnitudes of the observed period
changes are surprisingly large. Period changes should not be seen so easily,
yet they are, indicating that the evolution is a lot noisier than 
previously thought.

It has been argued (Lee 1991)
that 
for a given cluster, it is the {\it mean rate}
of period change that measures the evolutionary change, 
and that the {\it mean rate} of period change of RR Lyrae stars 
in globular clusters depends on horizontal branch morphology:
clusters with extremely blue horizontal branches have  
predicted  period increases occurring at a 
mean rate $\beta$=0.15 and up to 
0.5 d per million years, while 
for clusters with red HBs the rate of the predicted period decrease 
is smaller (Lee 1991). However, one should be aware that in presence
of noise even the mean rates of change could be misleading.

Observations do not give definitive results.
CC01 find that 30 of their M3 RR Lyrae variables lie above the ZAHB at
intensity-averaged magnitude brighter than 15.55. They argue whether 
these stars are evolved objects. 
Sixteen of these overluminous stars 
 have increasing periods, while 
nine have decreasing periods, and five have no previously determined 
period, thus  
tending to give some support the predictions of Lee (1991) models. 
However, of the stars lying 
above but closer to the ZAHB
(V intensity-averaged magnitudes between 15.65 and 15.72) for which 
evolutionary models would predict blueward evolution (i.e. decreasing periods) 
26 have increasing periods, 25 decreasing, two have 
constant period, and three have no previously determined period.

\subsubsection{PERIOD CHANGES IN DOUBLE-MODE RR LYRAE STARS}
Detection of changes in period in RR Lyrae stars usually require observations 
covering large time spans (of the order of decades), since 
variations are usually small and occur very slowly. However, 
changes in double-mode RR Lyrae stars seem to occur much faster 
than for single-mode stars, probably because these variables lie in the 
mode-switching region of the HB instability strip. 
Thus monitoring changes in period/mode of RRd's may require shorter
time spans.
Observational evidence exists that changes in 
double-mode RR Lyrae stars can occur on relatively short timescale: 
M68-V21 (Clement et al.\ 1993), M15-V31 (Purdue et al.\ 1995), 
and  M3-V79 (Clement et al.\ 1997). 

If an RRd star is evolving blueward, 
the amplitude of the first overtone should gain in strength relative to that 
of the fundamental. Vice versa, the strength of the 
fundamental mode oscillations should increase if the stars evolves redward.
Analysis of the various data sets available so far for some of the 
M3 RRd's (Clement et al.\ 1997) show that the first overtone
oscillations of both M3-V68 and M3-V79 have grown in strength since 1920s.
For V79, the change is more striking. The analysis shows also that
the fundamental period of V79 has decreased by a substantial amount in the 
last 35 years. Corwin et al.\ (1999) find a possible increase in the 
first overtone pulsation relative to the fundamental for M3-V99.

For the M3 RRd's 
for which so far there have been clear changes in the amplitude ratio,
  it is the strength of the 
first overtone
 pulsation that is increasing relative to the fundamental. 
Changes in the period and amplitude ratios of the double-mode RR Lyrae stars 
in the Oosterhoff type II cluster M15 are just the opposite than seen 
in the OoI cluster M3. This should
 indicate that the two clusters' double-mode variables 
  are evolving in different directions 
 on the horizontal branch, according to their different Oosterhoff types.
However, period and mode changes observed in the Oosterhoff type II cluster 
M68 do not seem to fit into the above framework.
 
According to Purdue et al.\ (1995), the ratio of the first overtone to 
fundamental
mode of pulsation has decreased in M15-V30 and possibly in M15-V53. 
Changes have occurred also in the amplitude ratio: 
the fundamental mode
oscillations of M15-V30 have increased in strength relative to the first overtone
between 1941 and 1991 with a rather abrupt change in the 1950s.
The same may be true of V53, though observational evidence is weaker.
M15-V26 does not appear to be a double-mode variable in 
1938, while it did show double-mode behavior with first overtone 
dominant mode in observations 
in 1951 (Jurcsik \& Barlai 1990). 
All these period/mode changes are consistent with redward evolution.

Clement et al.\ (1993) found that
the strength of the 
 secondary (fundamental) mode of pulsation of M68-V21 was varying during the 
interval of their observations. The star presented only a
primary first overtone period with no significant secondary pulsations in the
analysis of its full data set 
(1986-1988 and 1989-1991 interval), while when the 1986-1988 data 
were analyzed alone they revealed  
pulsation in both the fundamental and
first overtone modes, and that the amplitudes of the two modes were 
approximately 
equal. 
Walker (1994) found V21 to show double-mode behavior in observations taken 
in 1993.
Thus it seems that double-mode pulsation in V21 is a 
transient phenomenon.
M68-V33 also is reported to have changed from an RRd to
RRc in the time span going from 1950 to 1986 (Clement et al.\ 1993).  
Walker (1994) confirms that this star is no longer an RRd. 
Thus the situation with the M68 RRd's is controversial and does not
seem to be consistent with redward evolution.

 A situation similar to M15-V30 may have occurred to the field RRd star
AQ Leo, which has periods and period ratio similar to those of RRd stars
in Oosterhoff type II systems like M15. 
Jerzykiewicz, Schult, \& Wenzel (1982) found that the amplitude 
of the first overtone has decreased by an amount 
0.012$\pm$0.011, while the fundamental mode amplitude increased by 
0.017$\pm$0.011 in the time span from 1960 to 1974, with 
 an abrupt increase of the 
first overtone period in the early 1970s. However this variable is not of
much help in the present discussion on the connection between Oosterhoff types and 
direction of the HB evolution, since a field star cannot be 
associated with horizontal branch morphology, and so the directions of
period changes cannot be so easily predicted.
 
\subsubsection{MODE SWITCHING}
Corwin et al.\ (1999) found that M3-V166 has switched its dominant pulsation mode
from fundamental in 1992 to first overtone in 1993. 
In the present analysis we have found that also M3-V200 and M3-V251 have 
switched their dominant pulsation modes 
 in the same time span.
Mode switching is suggested as a possibility also for M15-V30 (Purdue et al.\ 1995) 
and for AQ Leo (Jerzykiewicz, Schult, \& Wenzel 1982) but none of these two
studies find direct evidence for the switch to have actually occurred in these 
stars.
Statistics on mode switching are still very poor. Indeed, M3 remains so far the only
stellar system where mode switching has been clearly observed. 
It is not clear whether this is an observational bias, and it is possible that 
further observations covering large time spans may reveal other mode switching 
RRd's among the double-mode RR Lyrae stars in M3 and other globular clusters.
Together with new, better time-resolved theoretical models for RRd stars,
the new data may help to understand whether mode switching is a natural phase in the
evolution of double-mode stars expected theoretically,
or it is, instead, just a further manifestation of the ``noise" 
that accompanies the HB evolution.

\subsubsection{THE BLAZHKO PHENOMENON}
Some connection very likely exists between period changes, 
switching of pulsation mode, double-mode pulsation and 
the Blazhko phenomenon (Blazhko 1907), 
a variation in both shape and amplitude of the light curve 
observed in about 30\% of known Galactic RR Lyrae
stars, that is superposed on the main period and which occurs with 
periodicities from a few days (e.g. 11 days, AH Cam, Smith et al.\ 1994) 
to a few hundreds of days (e.g. $\sim$ 530 d, RS Bootis,
Jones et al.\ 1988, and references therein).

Clementini et al.\ (1994) found that M4-V15, previously
claimed to be an ab-type RR Lyrae (Sturch 1977, Cacciari 
1979) in the photometric observations of 1986 showed a c-type light
curve, and in 1989 an ab-type light curve with smaller amplitude
and a different shape of the curve from previous studies. 
Clementini et al.\ (1994) suggested that V15 may be progressively
changing its mode of pulsation, going from ab to c-type, while 
evolving towards the blue part of the RR Lyrae instability strip
although they do not totally rule out that the star may be affected
by a strong Blazhko effect. 
Walker (1994) found that seven of the fundamental mode
RR Lyrae stars in M68, a cluster with a large 
number of RRd's,  show the Blazhko effect, and only
about half of the RR Lyrae stars in this cluster are singly periodic 
with stable light curves.
Clement et al.\ (1997) found that prior to 1962, M3-V79 was an RRab star with
an irregular light curve, a ``Blazhko" variable, later became a double-mode 
pulsator. They suggested that this supports the explanation of the Blazhko 
effect in terms of mixing of pulsational modes.
About 25\% of the M3 RR Lyrae stars in CC01 sample have been recognized to be 
affected 
by the Blazhko effect by Cacciari et al.\ (2003).

\subsubsection{PERIOD/MODE VARIATIONS AND THE DIRECTION OF THE HB EVOLUTION}
Based on previous observational evidence 
Clement et al.\ (1997) concluded that
changes in period and amplitude ratios of the double-mode
RR Lyrae stars in M3 and M15 indicate that  
while OoI clusters like M3  
are
evolving blueward, Oo II systems like M15 may be evolving redward.

However, variations observed in the M68 RRd's do not seem to be consistent with 
this interpretation.
Moreover, our new analysis shows that of the 3 newly discovered M3 double-mode
RR Lyrae,  
 M3-V200 and M3-V251 have both switched their 
dominant pulsation modes in the time span from 1992 to 1993, and  
while M3-V200 switched from fundamental to first overtone 
dominant mode similarly to M3-V166 (Corwin et al.\ 1999), 
in M3-V251 the mode switching has occurred in just the opposite way
(i.e. from first overtone to fundamental dominant pulsation mode).
This implies that if the evolutionary interpretation of
the mode switching is correct,  
both blueward and redward evolution
are occurring among the M3 double-mode RR Lyrae stars. 
Also, the observation of a switch in the dominant 
mode in less than 1 yr, as seen for V166, V200 and V251,  
is somehow in contrast with the small rate in period change predicted for
the Oo type I clusters by the evolutionary models.

We conclude that both extent and actual role of evolution on the period/mode 
variations of the RRd stars still remain unclear.
On the other hand, if
 evolution is not the culprit of the mode switching observed in the
M3 RRd's, as suggested by Purdue et al.\ (1995) in their concluding
remarks, one must invoke some unknown instability 
which may be responsible for the changes in period ratios, amplitude 
ratios and/or dominant pulsation modes observed in M3 and M15.

\section{SUMMARY AND CONCLUSIONS}
Image subtraction analysis of the photometric data by CC01 has lead to 
recovering 15 RR Lyrae stars listed in Bakos et al.\ (2000) catalogue, but 
not found by CC01, and has resulted in improved
periods for several other variables. We have identified three new double-mode RR Lyrae
stars in M3 (V13, V200, and V251). A rough estimate of their masses has been 
obtained based on B01 pulsational models and the Petersen diagram. We find that
both mass dispersion and strong evolutionary effects seem to be present among the
RRd's, and the RR Lyrae stars in general, in M3. Two of the newly discovered
RRd's have in fact period ratios as short as 0.738---0.739, and are well
separated from all known RRd's in the Petersen diagram, at positions implying
variations in mass by 0.1-0.2 M$_{\odot}$ and/or in the total heavy element content
by a factor 2-2.5, among the M3 giants.  
However, given the rather homogeneous [Fe/H] abundance, the constant $\alpha$-element 
enhancement, and the constant total [(C+N+O)/Fe] abundance derived for
the M3 stars from several independent spectroscopic studies, the latter
hypothesis seems rather unlikely. 

Three out of the 8 M3 double-mode pulsators (V68, V79, and possibly V99) 
show variations in length and amplitude of the two pulsation periods. 
Moreover, Corwin et al.\ (1999) and the 
present study have shown for the first time that the M3 RRd variables 
V166, V200, and 
V251 have 
switched their dominant pulsation modes in a very short time-span (about one 
year) thus suggesting that these stars are undergoing a rapid 
evolutionary phase, and giving support to the evolutionary interpretation of
the double-mode phenomenon. 

Clear evidence for evolutionary effects occurring among the 
single-mode
RR Lyrae stars in M3 has also been found recently by Cacciari et al.\ (2003) 
in their re-analysis of CC01 dataset.
They found that about 25\% of the M3 RR Lyrae stars 
in CC01 sample are affected by the Blazhko effect, and that 5\% are more 
evolved than the average luminosity level of the M3 RR Lyrae stars.

Since changes in the pulsation characteristics of the M3 RRd's seem to occur 
on a much shorter time-scale than for single-mode RR Lyrae stars, RRd's may be
a much more powerful tool to derive clues on the direction and rate of evolution
along the HB, through the monitoring of their fast changes in period and
pulsation modes. 

However, drawing firm conclusions on the actual direction of the evolution 
on the 
M3 HB on the basis of the present double-mode results seems premature here, 
given the small number of objects, the opposite 
behavior of V200 and V166 with respect to V251 (switching from fundamental to first overtone 
dominant modes, and vice versa, respectively), and the still rather
short time base-line of the present observations. 
In this respect we recall that our results do not well fit into the
scenario proposed by Clement et al.\ (1997) of the OoI clusters like
M3 evolving blueward, and the OoII clusters like M15 evolving 
redward, since both blueward and redward evolution
seems to occur among the M3 double-mode RR Lyrae stars.




A number of questions still remain open:
\begin{itemize}
\item [1] Is what we see in M3 a peculiarity of this cluster or are there other switching mode
RRd stars that still lie undetected in globular clusters and in the general field?


\item [2] Can the Petersen diagram be applied reliably to double-mode RR Lyrae stars undergoing rapid 
evolution to derive masses or other physical properties? Conversely, how does evolution affect the
straightforward relation thought to exist between period and density
(hence mass) via the Ritter and van-Albada and Baker equations? 
We know that evolutionary models are not adequate for interpreting the large
changes in period observed in the RR Lyrae stars and their rates.
Similarly, it may well be 
that the pulsational models and the ``theoretical'' Petersen diagram do 
not adequately predict masses for variable stars undergoing rapid 
evolutionary processes.

\item [3] Is the anomalous spread in mass of the M3 RRd's real? If it is 
real, what mechanisms are causing it: mass-transfer in binary systems, helium 
enhancement, or, less likely, varying $\alpha$-element enhancement among 
the M3 stars?
\end{itemize}

Additional data 
and continued monitoring of field and cluster RRd's over
long time spans (decades) are needed to reveal changes in period 
and amplitude ratios, and mode switching,
and to disentangle the role of evolution on the M3, M15 and M68 
single and double-mode RR Lyrae variables and on the HB stars 
in general. 

Elemental abundance analysis and dynamical studies 
are required to understand where the unusual masses of the M3 RRd's originated.

Radial velocity monitoring should also be undertaken to 
reveal possible binary systems and 
check whether 
mass transfer might be the cause of the low masses. 

Finally,  systematic searches with the image
subtraction technique extending to the crowded cluster cores should be performed to reveal 
whether further RRd's still lie undetected in other Galactic globular clusters. 

\section{Appendix: Notes on Individual Stars}

V13: This star is a candidate double-mode pulsator.

V29 and V155: In CC01 these two variables are interchanged as in Evstigneeva et al.\ (1994).

V44: This star has a very noisy light curve. Periods of about a half day and about one 
third day seem to phase the data reasonably well. Based on the shape of the light curve, 
we believe that it is and RRc variable and estimate the period at 0.33785. A period of 
about a half day was given in CC01 where it was listed as an RRab variable.

V180: This star was not found by CC01. In the web version of 
      Clement et al.\ (2001) ~~~Catalogue of ~~Variable ~~Stars ~~in 
      ~Galactic ~~Globular ~~Clusters, ~~available ~~at ~~   
      http://www.astro.utoronto.ca/people.html, the star is classified as 
      fundamental mode RR Lyrae, but no period is provided.
      The present data show that V180 is an ab-type RR Lyrae with period 0.61593 d. 
 
V200: This star is a candidate double-mode pulsator.

V234: This star was misidentified in CC01 as V164. Our period for this
      star is 0.04073 d shorter than in Strader et al.\ (2002).

V239: This star was mistakenly identified in CC01 as an RRc variable with 
      a period of 0.33343 d. This period is an alias of the correct period 
      0.66949 d. The shape of the light curve for V239 clearly identifies it 
      as an RRab. This mistake was noted by Strader et al.\ (2002).

V242: The period we derive for this
      star is 0.05805 d shorter than in Strader et al.\ (2002).

V245: This star was misidentified in CC01 as V198.

V250: The period we derive for this
      star is 0.03171 d longer than in Strader et al.\ (2002).

V251: This star is a double-mode pulsator.

V252: This star has a very unusual light curve (see Figure 15). The new period
      is 0.05091 longer than in CC01. 

V261: This star has very similar periods in CC01, the present paper and 
      Strader et al.\ (2002, P$\sim$0.44 d). However, it is classified 
      RRab by Strader et al.\
      and RRc by CC01. The shape and amplitude of the light 
      curve of V261 in Figure 1 of CC01 seem more consistent with 
      a first overtone pulsator.
        
V270: The period we derive for this
      star is 0.19637 d longer than in Strader et al.\ (2002), and in 
      good agreement with Clement et al.\ (2001) value of 0.69030 d.

\acknowledgments
We are grateful to Marcella Marconi for many helpful discussions and comments 
on the effects of $\alpha$-enhancement and evolution on the theoretical Petersen
diagram, to Carla Cacciari, Raffaele Gratton and Franca D'Antona 
for comments on the metallicity and helium
spreads in M3 and on the evolutionary effect in the M3 RR Lyrae stars, 
and to the
referee J. Nemec for his helpful comments.
BWC thanks the National Science Foundation for supporting this work through grant
AST-9988156 to the University of North Carolina.

\clearpage

\figcaption{1993 differential $B$ flux light curves for the 15 variable stars not found in 
Corwin \& Carney (2001). 
The data are for seven consecutive nights in 1993 and range from HJD 2449085 or 2449091. 
The order of the data is: filled squares, open squares, filled triangles, open triangles, 
filled circles, open circles, and crosses. 
 \label{fig1}}
\figcaption{Metallicity distribution of all known double-mode RR Lyrae stars. 
Metal abundances are on the ZW84 and/or Clementini et al.\ (1995) metallicity scales 
(see Table 3, 4 and 5).
These two scales are very similar. 
Only the 4 RRd's with known metal abundance are plotted of the MW field sample. Metallicities for the
other galaxies refer to the whole system and not to individual RRd's.
The open triangle 
with the arrow indicates the position in this diagram of the total number of the 
LMC field
RRd's (Soszy\'nski et al.\ 2003) assuming the average metal abundance of the 
LMC RR Lyrae in Clementini et al.\ (2003b)
transformed to the ZW84 scale. 
\label{fig2}}
\figcaption{RR Lyrae specific frequency (S$_{\rm RR}$; top panel) and horizontal-branch ratio
(HBR; bottom panel) of the Galactic globular clusters (open circles) as a function of metal abundance.
The 6 GGCs which host double-mode RR Lyrae are marked by filled circles.
Also shown are the 4 extragalactic
clusters which contain double-mode RR Lyrae stars (LMC: filled triangles, Fornax 3: filled
pentagon). ${\rm S_{RR}}$, HBR and [Fe/H] values 
are from Table 3.
 \label{fig3}}
\figcaption{Number of double-mode RR Lyrae stars in GCs versus: RR Lyrae 
specific frequency (${\rm S_{RR}}$, top panel),
horizontal-branch ratio (HBR, middle panel), and cluster central concentration (c, bottom 
panel). The double c denotes core-collapsed clusters.
Open and filled circles are used for the OoI and OoII Galactic clusters, respectively. 
Filled triangles are
the LMC clusters, the filled pentagon is Fornax 3. Cluster parameters are given 
in Table 3.
\label{fig4}}
\figcaption{Number of double-mode variables versus core radius (${\rm r_c}$, top 
panel), half-mass radius (${\rm r_h}$, middle panel), and tidal radius 
(${\rm r_t}$, bottom panel) of the cluster. 
Symbols are as in Figure 4. 
\label{fig5}}
\figcaption{Periodograms 
of the DAOPHOT $V$ data of V13 
obtained with GRATIS.
}
\figcaption{Light curves for V13 using the DAOPHOT V data. 
}
\figcaption{Light curves of V13 using the DAOPHOT $B$ data.\label{fig8}}
\figcaption{Periodograms for V200 using Alard B differential fluxes. 
}
\figcaption{Light curves for V200, 
using 
Alard B differential fluxes. 
\label{fig10}}
\figcaption{Periodograms for V251 using Alard B differential fluxes. 
\label{fig11}}
\figcaption{Light curves for V251 using the Alard B differential fluxes.
\label{fig12}}
\figcaption{Petersen diagram for all known double-mode RR Lyrae stars 
identified so far in various different stellar systems
for which periods are available in the literature. Filled circles 
identify the
three new candidate RRd's reported in this paper, filled square mark the 
previously known M3 RRd's. Lines show the pulsational models
by Bono et al.\ (1996), which correspond to Z=0.0001, and M/M$_{\odot}$=0.65, 
0.75, and 0.80, and their extension to larger mass and metallicity ranges
(M/M$_{\odot}$=0.85, and Z=0.0002, 0.0004, 0.0006, 0.0008) obtained by B01.
\label{fig13}}
\figcaption{Enlargement of Figure 13 in the region relevant for the M3 RRd's. Lines
corresponding to the different pulsational models are labelled by mass and luminosity
level (in solar units).\label{fig14}}
\figcaption{1993 differential $B$ flux light curve for V252. Symbols as in 
Figure 1. Data are folded according ot the new period given in Table 2.\label{fig15}}

\clearpage

\begin{table}[ht]
\caption{Ephemerides for the variables recovered in the present analysis}
\vspace{0.5cm}
\begin{tabular}{lclcc}
\tableline
\tableline
\multicolumn{1}{l}{Variable}  & 
\multicolumn{1}{c}{Period}    & 
\multicolumn{1}{c}{Epoch}     & 
\multicolumn{1}{c}{Period}    & 
\multicolumn{1}{c}{Type}      \\

\multicolumn{1}{l}{}  & 
\multicolumn{1}{c}{(1)}    & 
\multicolumn{1}{c}{(1)}     & 
\multicolumn{1}{c}{(2)}    & 
\multicolumn{1}{c}{(1)}      \\

\tableline
V13     &0.47951/0.35382 &2448756.820 & --- &  candidate RRd\\
V180    &0.61593 &2449091.787 & --- &  RRab \\
V200    &0.48674/0.35982 &2449090.842 & 0.5292& candidate RRd\\
V210    &0.35409 &2449087.972 & --- & RRc~\\
V242    &0.59325 &2449086.712 & 0.6513& RRab\\
V244    &0.53786 &2449090.809 & --- & RRab\\
V249    &0.53301 &2449087.724 & --- & RRab\\
V250    &0.59031 &2448753.775 & 0.5586 & RRab\\
V251    &0.47423/0.35384 &2448756.820 & --- & candidate RRd\\
V253    &0.33161 &2449088.936 & 0.3328 & RRc~\\
V254    &0.60455 &2449089.769 & 0.6056 & RRab\\
V255    &0.57264 &2449091.849 & --- & RRab\\
V257    &0.60095 &2449085.880 & 0.6019 & RRab\\
V264    &0.35649 &2448755.093 & 0.3565 & RRc~\\
V270    &0.69017 &2448754.879 & 0.4938 & RRab\\
V271    &0.63043 &2449088.870 & 0.6329 & RRab\\
\tableline
\end{tabular}
\medskip

(1) Period, epoch of maximum light and type classification found by our analysis.

(2) Period from Strader et al.\ (2002).
\end{table}

\begin{deluxetable}{lccll}
\footnotesize
\tablecaption{Variables with revised periods \label{tbl-2}}
\tablewidth{0pt}
\tablehead{
\colhead{Variable} & \colhead{New Period} & \colhead{CC01} &  
\colhead{CC01 diff.} &\colhead{Type} }
\startdata
V8      &0.63800 &0.63916  &-0.00116 &RRab  \\ 
V44	&0.33785 &0.50635  &-0.16850 &RRc, noisy curve \\
V115	&0.51258 &0.51335  &-0.00077 &RRab \\
V122    &0.51609 &0.50603  &0.01006  &RRab  \\  
V148    &0.46729 &0.46597  &0.00132  &RRab  \\
V149    &0.54815 &0.54996  &-0.00181 &RRab  \\
V157	&0.54282 &0.54187  &0.00095  &RRab  \\
V159	&0.53326 &0.53370  &-0.00044 &RRab  \\
V161    &0.52164 &0.51444  &0.00720  &RRab  \\
V165	&0.48363 &0.48504  &-0.00141 &RRab  \\
V168	&0.27597 &0.27643  &-0.00046 &RRc  \\
V170    &0.43226 &0.43569  &-0.00343 &RRc \\
V174	&0.58703 &0.59450  &-0.00747 &RRab  \\
V175    &0.56970 &0.56584  &0.00386  &RRab  \\
V181(H903)    &0.66384 &0.66790  &-0.00406 &RRab  \\
V184    &0.53213 &0.52958  &0.00255  &RRab  \\
V189    &0.61294 &0.61869  &-0.00575 &RRab  \\
V191	&0.52003 &0.51921  &0.00082  &RRab  \\
V192    &0.49790 &0.48145  &0.01645  &RRab  \\
V193    &0.74784 &0.73285  &0.01499  &RRab  \\
V201    &0.54141 &0.53963  &0.00178  &RRab  \\
V207    &0.34458 &0.34494  &-0.00036 &RRc  \\
V208	&0.33802 &0.33735  &0.00067  &RRc  \\
V209    &0.34939 &0.34718  &0.00221  &RRc  \\
V214	&0.54045 &0.53949  &0.00096  &RRab, Blazhko  \\
V215	&0.52710 &0.53308  &-0.00598 &RRab  \\
V218    &0.54487 &0.54308  &0.00179  &RRab  \\
V219    &0.61366 &0.61139  &0.00227  &RRab  \\
V220    &0.60011 &0.59585  &0.00426  &RRab \\
V222	&0.50151 &0.50065  &0.00086  &RRab \\
V226	&0.48843 &0.48769  &0.00074  &RRab \\
V234(V164)   &0.50877 &0.50959 &-0.00082&RRab   \\
V235    &0.76519 &0.76175  &0.00344 &RRab \\
V239(X13)    &0.66949 &0.33343 &0.33606 &RRab     \\
V240(X14)    &0.27601 &0.27647 &-0.00046 &RRc   \\
V241(X17)    &0.59408 &0.59617 &-0.00209 &RRab   \\
V243(X22)    &0.62991 &0.63220 &-0.00229 &RRab   \\
V245(V198)   &0.28403 &0.28305 &0.00098  &RRc  \\
V246(X30)    &0.33914 &0.33845 &0.00069  &RRc  \\
V248(X36)    &0.51134 &0.51065 &0.00069  &RRab   \\
V252(KG4)    &0.50155 &0.45064 &0.05091  &?  \\
V256(KG11)   &0.31836 &0.31564 &0.00272  &RRc  \\
V259(KG15)   &0.33287 &0.33352 &-0.00065 &RRc  \\
V261(H812)   &0.44414 &0.44467 &-0.00053 &RRc  \\
\enddata
\end{deluxetable}

\begin{table}[ht]
\caption{Informations on the RRd variables identified in globular clusters}
\vspace{0.5cm}
\tiny
\begin{tabular}{lrrccrccccccc}
\tableline
\tableline
\multicolumn{1}{c}{ID-Name}&\multicolumn{1}{c}{${\rm N_{RR}}$}&
 \multicolumn{1}{c}{${\rm N_{RRd}}$} &
\multicolumn{1}{c}{[Fe/H]} & \multicolumn{1}{c}{Oo} &
\multicolumn{1}{c}{${\rm S_{RR}}$} &\multicolumn{1}{c}{HBR} & 
\multicolumn{1}{c}{c} & \multicolumn{1}{c}{${\rm r_{c}(\prime)}$} & 
\multicolumn{1}{c}{${\rm r_{h}(\prime)}$} &\multicolumn{1}{c}{${\rm r_{t}(\prime)}$} &
\multicolumn{1}{c}{${\rm <M/M_{\odot}>}$} & 
\multicolumn{1}{c}{${\rm \sigma_{M}}$}\\

\multicolumn{1}{c}{}&\multicolumn{1}{c}{(a)} &\multicolumn{1}{c}{} &
\multicolumn{1}{c}{(b)} & \multicolumn{1}{c}{(c)} &
\multicolumn{1}{c}{(d)} &\multicolumn{1}{c}{(e)} & 
\multicolumn{1}{c}{(f)} &
\multicolumn{1}{c}{(f)} &
\multicolumn{1}{c}{(f)} & 
\multicolumn{1}{c}{(f)} &
\multicolumn{1}{c}{(g)} &
\multicolumn{1}{c}{(g)} \\

\tableline
\multicolumn{13}{c}{Milky Way}\\

 IC4499      &97  & 16 &  -1.50 & I &113.4  & 0.11         & 1.11       &0.96     &1.50   &12.35    &  0.853  & 0.023\\
 NGC5272-M3  &182 & 5+3 & -1.66 & I & 49.0  & 0.08         & 1.84       &0.55     &1.12   &38.19    &  0.779  & 0.063\\
 NGC4590-M68 &42  & 12 & -2.09 & II & 48.3  & 0.17         & 1.64       &0.69     &1.55   &30.34    &  0.755  & 0.019\\
 NGC2419     &31  &  1 & -2.10 & II &  4.6  & 0.86         & 1.40       &0.35     &0.73   & 8.74    &  0.793  &    $-$\\
 NGC7078-M15 &88  & 14 & -2.15 & II & 18.9  & 0.67         & 2.50       &0.07     &1.06   &21.50    &  0.785  & 0.036\\
 NGC6426     &14  &  1 & -2.20 & II & 25.3  & 0.58         & 1.70       &0.26     &0.96   &13.23    &  0.814  &    $-$\\
& & & & & & & & & & & &\\
\multicolumn{13}{c}{Large Magellanic Cloud}\\
Reticulum   &32 &  4  & -1.71 & I   &132.2  &-0.04$\pm$0.05&1.0         &1.0           &1.58   &10.0     &  $-$     &$-$ \\
NGC1835     &84 &  6  & -1.79 & I/II& 18.2  &$-$           &1.5         &0.10          &0.32   & 3.16    &  $-$     &$-$ \\
NGC2019     &41 &  2  & -1.81 & $-$ & 27.3  &$-$           &1.6$\pm$0.3 &0.07$\pm$0.03 &0.24   & 2.63    &  $-$     & $-$   \\
& & & & & & & & & & & &\\
\multicolumn{13}{c}{Fornax}\\
Fornax 3    &99 &  8  & -1.91 &II/(I)&90.2  &0.40$\pm$0.05 &1.28     &$0.07\pm 0.02$      &0.16 & 1.28$\pm 0.08$  &$-$ &$-$\\
            &   &     &       &      &      &              &1.83     &$0.02\pm 0.01$      &0.12 & 1.58$\pm 0.37$  &    &   \\
\tableline
\end{tabular}
\medskip

$^a$ N$_{\rm RR}$ values are from Clement et al.\ (2001) for
the MW clusters, from Walker (1992) and Ripepi et al.\ (2003) for
the Reticulum, from Soszy\'nski et al.\ (2003) for NGC1835 and
NGC 2019, from Mackey \& Gilmore (2003c) for Fornax 3.

$^b$ Metall abundances for the MW clusters are from ZW84.
Metallicities for the LMC clusters are taken from Suntzeff
et al.\ (1992), and from Clementini et al.\ (2003a)
for Fornax 3. They are on the ZW84 metallicity scale.

$^c$ The Oosterhoff type classification (Oo) for the Galactic 
clusters is taken from Clement et al.\ (1991), 
for the Reticulum cluster is from Walker (1992) and 
Ripepi et al.\ (2003), for NGC 1835 is from
Soszy\'nski et al.\ (2003), and 
for Fornax 3 from Clementini et al.\ 
(2003a).

$^d$ The RR Lyrae specific frequency for the Milky Way clusters 
is from the 2003 update of Harris (1996) web catalogue on GGCs.
For the LMC clusters S$_{\rm RR}$ was computed from the N$_{\rm RR}$ values in column
2 adopting $M_V$ values from Suntzeff et al.\ (1992), for
Fornax 3 is from Mackey \& Gilmore (2203c).

$^e$ The HBR value of the MW clusters is from Harris 
catalogue, for Reticulum is from Walker (1992), for Fornax 3
is from Mackey \& Gilmore (2203c). 

$^f$ 
Structural parameters  for the MW clusters are from
Harris catalogue.
c, r$_c$, r$_t$ values for the LMC clusters are from Suntzeff et al.\ (1992),
for Fornax 3 are from  Demers et al.\ (1994) and Webbink (1985;  
see also Mackey \& Gilmore 2003a,b, for  
new estimates of the core radius of NGC 1835, NGC 2019, and
Fornax 3).  
Half mass radii (r$_{\rm h}$) for the extragalactic clusters
have been computed according to the definition given in 
Harris catalogue ($log{\rm (r_h/r_c)} = 0.6 \times$c $-$0.4).

$^g$ Average masses and ${\rm \sigma_{M}}$ values of the MW globular cluster
RRd's have been computed from the values published in B01. 
In this paper the interested reader can find individual pulsational
masses for the Galactic cluster RRd's and for some of the field RRd's in the 
MW and in the LMC.  
\end{table}

\begin{table}[ht]
\caption{Informations on the field RRd variables 
found in Local Group galaxies}
\vspace{0.5cm}
\footnotesize
\begin{tabular}{lrcccccc}
\tableline
\tableline
\multicolumn{1}{c}{ID-Name}&
\multicolumn{1}{c}{~~~${\rm N_{RRd}}$} &
\multicolumn{1}{c}{[Fe/H]$^a$} & 
\multicolumn{1}{c}{Oo}&
\multicolumn{1}{c}{r$_c^b$}&
\multicolumn{1}{c}{r$_c^b$}&
\multicolumn{1}{c}{r$_t^b$}&
\multicolumn{1}{c}{Ref. (for [Fe/H])}\\

\multicolumn{1}{c}{}&
\multicolumn{1}{c}{} &
\multicolumn{1}{c}{} & 
\multicolumn{1}{c}{} & 
\multicolumn{1}{c}{(arcmin)} & 
\multicolumn{1}{c}{(pc)} & 
\multicolumn{1}{c}{(arcmin)} & 
\multicolumn{1}{c}{}\\

\tableline
Milky Way     &~22$^c$ & --   & I$^d$ & -- & -- & -- & --\\ 

 LMC          &230  & -1.54 &I/II& -- & -- & -- &1\\
\medskip
 SMC          &57   & -1.7~~  &I/II& -- & -- & -- & 2\\
\medskip
 Sagittarius  &40   & -1.6~~  &~~I/(II)& -- & 550 & $>10 \deg$ &3\\
 \medskip
 Sculptor     & 18  & -1.80 &I/II& 5.8$\pm 1.6$& 110& 76.5$\pm 5.0$ &4\\
\medskip
 Fornax       & 12 &  -1.77 &I/II&13.8$\pm 0.8$& 460 & 71$\pm 4$ &5\\
\medskip
 Draco        & 10  & -2.00 &II/I& 9.0$\pm 0.7$& 180& 28.3$\pm 2.4$ &4\\
\medskip
 Carina       &  6 &  -2.1~~ &II/I& 8.8$\pm 1.2$& 210&  28.8$\pm 3.6$ &6\\
 
\tableline
\end{tabular}
\medskip

$^a$Metall abundances are for the whole
system and not for individual RRd's,  
they are on or were
transformed to the ZW84 metallicity scale. 
               

$^b$Structural parameters for the dwarf galaxies are
taken from Mateo (1998).

$^c$The number of MW RRd's includes 6 previously 
knonw objects (see Clementini et al.\ 2000), 3 new
discoveries in the galactic bulge (Mizerski 2003), and
13 RRd's lying along the line of sight to Sagittarius (Cseresnjes 2001).

$^d$The Oosterhoff type classification for the Milky Way refers only to 
variables in the Galactic Center (Cseresnjes 2001).

References.- (1) Clementini et al.\ (2003b) transformed to the ZW84; 
(2) Smith et al.\ (1992); (3) Cseresnjes (2001);
(4) Mateo (1998); (5)  Clementini et al.\ (2003a);
(6) Smecker-Hane et al.\ (1999).

\end{table}

\begin{table}[ht]
\caption{Individual metallicities of field RRd variables}
\vspace{0.5cm}
\footnotesize
\begin{tabular}{lcc}
\tableline
\tableline
\multicolumn{1}{c}{ID-Name$^a$}&
\multicolumn{1}{c}{[Fe/H]$^b$} & 
\multicolumn{1}{c}{Ref.} \\


\tableline
 MW-RRVIII-58   & -1.75$\pm$0.20    &1\\
 MW-AQLeo       & -1.81$\pm$0.20    &1\\
 MW-RRVIII-10   & -1.86$\pm$0.20    &1\\
\medskip
 MW-CUCom       & -2.38$\pm$0.20    &2\\
 LMC-CA48 - 06.6811.651   & -1.38$\pm$0.13  &3\\
 LMC-CB61 - 13.5958.518   & -1.23$\pm$0.20  &1\\
 LMC-CB49 - 13.5836.525   & -1.71$\pm$0.13  &3\\
 LMC-CB45 - 13.6080.591   & -1.70$\pm$0.10  &3\\
 LMC-CA02 - 13.6691.4052  & -1.74$\pm$0.20   &1\\
 ~~~~~~~~~~~~~~~ - 06.6691.1003  &                 &\\
 LMC-CA67 - 06.6810.428   & -2.07$\pm$0.13  &3\\
 LMC-CB59 - 13.5838.497   & -1.54$\pm$0.09  &3\\
 LMC-CA05 - 13.7054.2970  & -1.60$\pm$0.27  &3\\
 LMC-CA12 - 06.6933.939   & -1.42$\pm$0.14  &3\\
\tableline
\end{tabular}
\medskip
               
$^a$For the LMC variables we list both B01 and Alcock et 
al. (1997) identification numbers.
\par\indent
$^b$Metall abundances are either on the ZW84 and/or 
on Clementini et al.\ (1995) metallicity scales. These two scales 
are very similar. Metallicities for 7 of the LMC RRd's were taken
from Gratton et al.\ (2003, in preparation) and transformed to the ZW84 scale according to the procedure
described in Clementini et al.\ (2003b).
\par\indent
References.- (1) B01; (2) Clementini et al.\ (2000); (3) Gratton et al.\ 
(2003, transformed
to the ZW84 metallicity scale).

\end{table}

\begin{table}[ht]
\caption{Parameters of the RRd variables in M3}
\vspace{0.5 cm}
\begin{tabular}{llllccc}
\tableline
\tableline
\multicolumn{1}{l}{Variable}        & \multicolumn{1}{c}{${\rm P_0}$}         & 
\multicolumn{1}{c}{${\rm P_1/P_0}$} & \multicolumn{1}{c}{${\rm M/M_{\odot}}$} & 
\multicolumn{1}{c}{$< V_{int}>$} & 
\multicolumn{1}{c}{$< B_{int}> - < V_{int}>$} &\multicolumn{1}{c}{$r$} \\

\multicolumn{1}{l}{}        & \multicolumn{1}{l}{}         & 
\multicolumn{1}{l}{} & \multicolumn{1}{l}{} & 
\multicolumn{1}{c}{(1)} & 
\multicolumn{1}{c}{(1)} & \multicolumn{1}{c}{(arcsec)} \\
\tableline
V68     & 0.479 & 0.7432 & 0.68$\pm$0.11    &15.642&0.288&186.9\\
V79     & 0.4797& 0.7453 & 0.77$\pm$0.14    &15.658&0.278&362.7\\
V87     & 0.48  & 0.7458 & 0.80$\pm$0.15    &15.578&0.274&134.4\\
V166    & 0.482 & 0.7459 & 0.80$\pm$0.15    &15.700&0.289&~92.8\\
V99     & 0.4835& 0.7468 & 0.85$\pm$0.16    &15.602&0.276&210.0\\
V251    & 0.47423& 0.7461 &~~~$\sim$ 0.8     & $-$ & $-$&~~3.7\\ 
V200    & 0.4867 & 0.739  &~~ $\sim$ 0.65     & $-$ & $-$&~32.0\\
V13     & 0.47951& 0.7379 & $\sim$ 0.55/0.65&15.612&0.270&129.1\\
\tableline
\end{tabular}
\medskip

(1) Intensity-averaged magnitudes and colors are taken from 
CC01, they are before pre-whitening. The values after pre-whitening derived for 
V13 in the present analysis 
are 15.662 and 0.268, respectively.

\end{table}


\clearpage
\begin{references}
\reference{} Alard, C. 2000, A$\&$AS, 144, 235
\reference{} Alard, C. \& Lupton, R. H. 1998, ApJ, 503, 325
\reference{} Alcock, C. et al.\ (The MACHO Collaboration) 1996, \aj, 111, 1146
\reference{} Alcock, C. et al.\ (The MACHO Collaboration) 1997, ApJ, 482, 89
\reference{} Alcock, C. et al.\ (The MACHO Collaboration) 2000, ApJ, 542, 257
\reference{} Armosky, B.J., Sneden, C., Langer, G.E., \& Kraft, R.P. 1994, 
 AJ, 108, 1364
\reference{} Baker, R. H. \& Baker, H. V. 1956, \aj, 61, 283
\reference{} Bakos, B. A., Benko, J. M., \& Jurcsik, J. 2000, Acta Astron., 50, 221
\reference{} Barning, F.J.M. 1963, \bain, 17, 22
\reference{} Blazhko, S. 1907, Astron. Nachr., 175, 325
\reference{} Bono, G., Caputo, F., Castellani, V., \& Marconi, M. 1996, ApJ, 471, L33
\reference{} Bragaglia, A., Gratton, R.G., Carretta, E., Clementini, G., Di Fabrizio, L., \&
             Marconi, M. 2001, \aj, 122, 207 (B01)
\reference{} Buonanno, R., Corsi, C. E., Buzzoni, A., Cacciari, C., Ferraro, F. R., \& 
             Fusi Pecci, F. 1994, A\&A, 290, 69
\reference{} Buonanno, R., Corsi, C. E., \& 
             Fusi Pecci, F. 1981, MNRAS, 196, 435
\reference{} Cacciari, C. 1979, \aj, 84, 1542
\reference{} Cacciari, C., Corwin, T.M., \& Carney, B.W., 2003, in preparation
\reference{} Caputo, F., Santolamazza, P., Marconi, M., 1998, MNRAS, 293, 364
\reference{} Carney, B. W. 1996, PASP, 108, 900
\reference{} Carretta, E., Cacciari, C., Ferraro, F. R., Fusi Pecci, F. \& Tessicini, 
             G. 1998, MNRAS, 298, 1005
\reference{} Carretta, E., \& Gratton, R.G. 1997, A$\&$AS, 121, 95
\reference{} Clement, C.M., Ferance, S., \& Simon, N. 1993, \apj, 412, 183
\reference{} Clement, C.M., Hilditch, R. D., Kaluzny, J., \& Rucinski, S. M. 1997, ApJ, 489, L55
\reference{} Clement, C.M., Kinman, T. D., \& Suntzeff, N. B. 1991, ApJ, 372, 273
\reference{} Clement, C.M., et al.\ 2001, \aj, 122, 2587
\reference{} Clementini, G., Carretta, E., Gratton, R., Merighi, R., Mould, 
             J.R., McCarthy, J.K. 1995, \aj, 110, 2319
\reference{} Clementini, G., et al.\ 2000, \aj, 120, 2054
\reference{} Clementini, G., et al.\ 2003a, in Variable stars in the Local Group, eds.
             D. Kurtz \& K. Pollard, ASP Conf. Series, in press  
\reference{} Clementini, G., Gratton, R.G., Bragaglia, A., Carretta, E., Di Fabrizio, L., 
             \& Maio, M. 2003b, AJ, 125, 1309 (astro-ph/0007471)
\reference{} Clementini, G., Merighi, R., Pasquini, L., Cacciari, C., \&
             Gouiffes, C. 1994, MNRAS, 267, 83             
\reference{} Corwin, T. M., Carney, B. W., \& Allen, D. M. 1999, \aj, 117, 1332
\reference{} Corwin, T. M. \& Carney, B. W. 2001, \aj, 122, 3183 (CC01)
\reference{} Cox, A. N., Hodson, S. W., \& Clancy, S. P. 1983, ApJ, 266, 94
\reference{} Cseresnjes, P. 2001, A\&A, 375, 909
\reference{} Dall'Ora, M., et al., 2003, \aj, in press (astro-ph/0302418)
\reference{} D'Antona, F., Caloi, V., Montalban, J., Ventura, P., \& Gratton, R. 2002,
             A\&A, 395, 69
\reference{} Demers, S., Irwin, M.J., Kunkel, W.E. 1994, \aj, 108, 1648
\reference{} Evstigneeva, N. M.,Samus, N. N., Tsvetdova, T. M. \& Shokin, Y. A. 1994, 
             Pisma Astron. Zh., 20, 693
\reference{} Ferraro, F. R., Carretta, E., Corsi, C. E., Fusi Pecci, F., Cacciari, C., 
             Buonanno, R, Paltrinieri, B., \& Hamilton, D. 1997a, A\&A, 320, 757
\reference{} Ferraro, F. R., Paltrinieri, B., Fusi Pecci, F., Cacciari, C., Dorman, B., 
             \& Rood, R. T. 1997b, ApJ, 484, L145
\reference{} Ferraro, F. R., et al.\ 1997c, A\&A, 324, 915
\reference{} Garcia-Melendo, E., \& Clement, C.M. 1997, \aj, 114, 1190
\reference{} Goranskij, V. P. 1981, Inf. Bull. Variable Stars, 2007
\reference{} Gratton, R.G., Bragaglia, A., Clementini, G., Carretta, E., Di Fabrizio, 
             L., Maio, M., Taribello, E. 2003, in preparation
\reference{} Gunn, J.E., \& Griffin, R.F. 1979, \aj, 84, 752
\reference{} Harris, W. E. 1996, \aj, 112, 1487
\reference{} Hut, P. et al.\ 1992, PASP, 104, 981 
\reference{} Iben, I, Jr., \& Rood, R.T. 1970, ApJ, 161, 587           
\reference{} Jerzykiewicz, M., \& Wenzel, W. 1977, Acta Astron., 27, 35             
\reference{} Jerzykiewicz, M., Schult, R. H., \& Wenzel, W. 1982, Acta Astron., 32, 357 
\reference{} Jones, R.V., Carney, B.W., \& Latham, D.W. 1988, \apj, 332, 206
\reference{} Jurcsik, J., \& Barlai, K., 1990, in Confrontation Between Stellar Pulsation and
             Evolution, eds. C. Cacciari \& G. Clementini, ASP Conf. Series, Vol. 11, p. 112          
\reference{} Kaluzny, J., Hilditch, R. W., Clement, C., \& Rucinski, S. M. 1998, MNRAS, 296,347
\reference{} Kaluzny, J., Kubiak, M., Szymanski, M., Udalski, A., Krzeminski, W., Mateo, 
             M. 1995, A\&AS, 112, 407
\reference{} Kovacs, G. 2001a, A\&A, 375, 469
\reference{} Kovacs, G. 2001b, in Stellar Pulsation-Nonlinear Studies, eds. M. Takeuti \& 
             D. D. Sasselov, Astrophysics and space science library, Vol. 257, (Kluwer Academic Publishers), 61
\reference{} Kraft, R. P. \& Ivans, I.I. 2003, PASP, 115, 143
\reference{} Kraft, R. P., Peterson, R.C., Guhathakurta, P., Sneden, C., 
             Fulbright, J.P., \& Langer, G. E. 1999, ApJ, 518, L53
\reference{} Kraft, R. P., Sneden, C., Langer, G. E., \& Prosser, C.F. 1992, AJ 104, 645
\reference{} Kraft, R. P., Sneden, C., Langer, G. E., Stetrone, M. D., \& Bolte, M. 1995, 
             \aj, 109, 2586
\reference{} Lee Y.-W. 1991, \apj, 367, 524  
\reference{} Lee S.-G. 1999, \aj, 118, 920  
\reference{} Lee Y.-W., Demarque P., \& Zinn R. 1990, \apj, 350, 155
\reference{} Lomb, N.R. 1976, \apss, 39, 447
\reference{} Mackey, A.D., \& Gilmore, G.F. 2003a, MNRAS, 338, 85
\reference{} Mackey, A.D., \& Gilmore, G.F. 2003b, MNRAS, 340, 175
\reference{} Mackey, A.D., \& Gilmore, G.F. 2003c, astro-ph/0307275
\reference{} Mateo, M. 1998, ARAA, 36, 435
\reference{} Mizerski, T. 2003, in Variable stars in the Local Group, eds.
             D. Kurtz \& K. Pollard, ASP Conf. Series, in press  
\reference{} Moskalik, P., \& Poretti, E. 2003, A\&A, 398, 213
\reference{} Nemec, J. M. 1985a, \aj, 90, 204 
\reference{} Nemec, J. M. 1985b, \aj, 90, 240
\reference{} Nemec, J. M., \& Clement, C.M. 1989, \aj, 98, 860
\reference{} Nemec, J. M., Nemec, A. F. L. \& Norris, J. 1986, AJ, 92, 358
\reference{} Norris, J., Cottrell, P.L., Freeman, K.C., \& Da Costa, G.S. 1981,
             \apj, 244, 205
\reference{} Norris, J., \& Smith, G.H. 1984, \apj, 287, 255
\reference{} Oosterhoff, P. 1939, Observatory, 62, 104
\reference{} Petersen, J.O. 1973, A\&A, 27, 89
\reference{} Pilachowski, C. A., \& Sneden, C. 2001, AAS 199, 137.08
\reference{} Popielski, B. L., Dziembowski, W. A., \& Cassisi, S. 2000, Acta Astron., 50.491
\reference{} Purdue, P., Silbermann, N.A., Gay, P., \& Smith, H.A. 1995, \aj, 110, 1712
\reference{} Pryor, C.P., Latham, D.W., \& Hazen, M.L. 1988, \aj, 96, 123
\reference{} Renzini, A. 1977, in Advanced Stages in Stellar Evolution, 7th Course of
 the Swiss Society of Astronomy and Astrophysics, Saas-Fee, p. 149, eds. P. Bouvier \&
 A. Maeder, Geneva Observatory, Sauverny 
\reference{} Ripepi, V., et al.\ 2003, in Asteroseismology and Stellar Evolution, 
             JENAM-2003 Symp., Budapest, Hungary, in press
\reference{} Roberts, M. S. \& Sandage, A. 1955, \aj, 60, 185
\reference{} Rood, R. T., Carretta, E., Paltrinieri, B., Ferraro, F. R., Fusi Pecci, 
             F., Dorman, B., Chieffi, A., Straniero, O., \& Buonanno, R. 1999, ApJ, 523, 752
\reference{} Rood, R. T., \& Crocker, D. 1985, in Production and Distribution of
             C,N,O Elements, eds. J. Danziger, F. Matteucci, K. Kjar (ESO, 
             Garching bei Munchen), p. 61
\reference{} Salaris, M., \& Cassisi, S. 1996, A\&A, 305, 858
\reference{} Sandage, A. 1953, \aj, 58, 61
\reference{} Sandage A. 1959, ApJ, 129, 596
\reference{} Sandage, A., Katem, B., \& Sandage, M. 1981, \apjs, 46, 41
\reference{} Scargle, J. D. 1992, ApJ, 263, 835
\reference{} Smecker-Hane, T.A., Mandushev, G.I., Hesser, J.E., Stetson, P.B., 
             Da Costa, G.S., \& Hatzidimitriou, D. 1999, in Spectrophotometric 
             Dating of Stars and Galaxies, eds. I. Hubeny, S. Heap \& R. Cornett, ASP 
             Conf. Series, Vol.192, p. 159 (astro-ph/9910212)
\reference{} Smith, H. A. 1995, RR Lyrae Stars (Cambridge: Cambridge Univ. Press)
\reference{} Smith, H. A., Matthews, J.M., Lee, K.M., Williams, J., Silbermann, N.A., 
             \& Bolte, M., 1994, \aj, 107, 679
\reference{} Smith, G.H., Shetrone, M.D., Bell, R.A., Churchill, C.W., \& 
             Briley, M.M. 1996, \aj, 112, 1511
\reference{} Smith, H. A., Silbermann, N., Baird, S., \& Graham, J. 1992, \aj, 104, 1430
\reference{} Soszy\'nski, I., et al.\ 2002, Acta Astronomica, Vol. 52, 369
\reference{} Soszy\'nski, I., et al.\ 2003, astro-ph/0306041
\reference{} Stellingwerf, R. F. 1978, ApJ, 224, 953
\reference{} Strader, J., Everitt, H. O., \& Danford, S. 2002, MNRAS, in press (astro-ph/0203135).
\reference{} Sturch, C.R. 1977, PASP, 89, 349
\reference{} Suntzeff, N.B. 1981, ApJS, 47, 1
\reference{} Suntzeff, N.B. 1993, in The Globular Cluster-Galaxy Connection, eds. G.H. Smith \&
             J.P. Brodie, ASP Conf. Series, Vol. 48, p. 14
\reference{} Suntzeff, N.B., Schommer, R.A., Olszewki, E.W., \& Walker, A. 1992, \aj, 104, 1743
\reference{} Sweigart A.V., \& Renzini A. 1979, A\&A, 71, 66
\reference{} Walker, A. R. 1992, \aj, 103, 1166
\reference{} Walker, A. R. 1994, \aj, 108, 555
\reference{} Walker, A. R., \& Nemec, J. M. 1996, \aj, 112, 2026
\reference{} Webbink, R.F. 1985, in Dynamics of Star Clusters, eds. J. 
             Goodman \& P. Hut, 
             IAU Symp. 113, p. 541, Kluwer, Dordrecht
\reference{} van Albada, T.S.,\& Baker, N. 1971, \apj, 169, 311
\reference{} VandenBerg, D. 1992, \apj, 391, 685
\reference{} Zinn, R. \& West, M. J. 1984, ApJS, 55, 45 (ZW84)
\end{references}
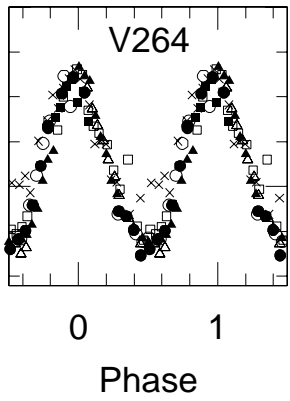
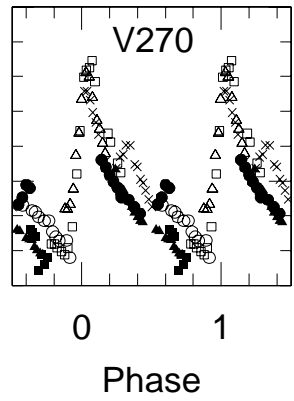
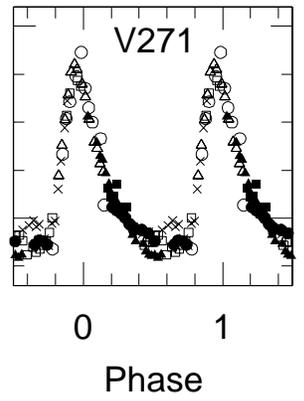
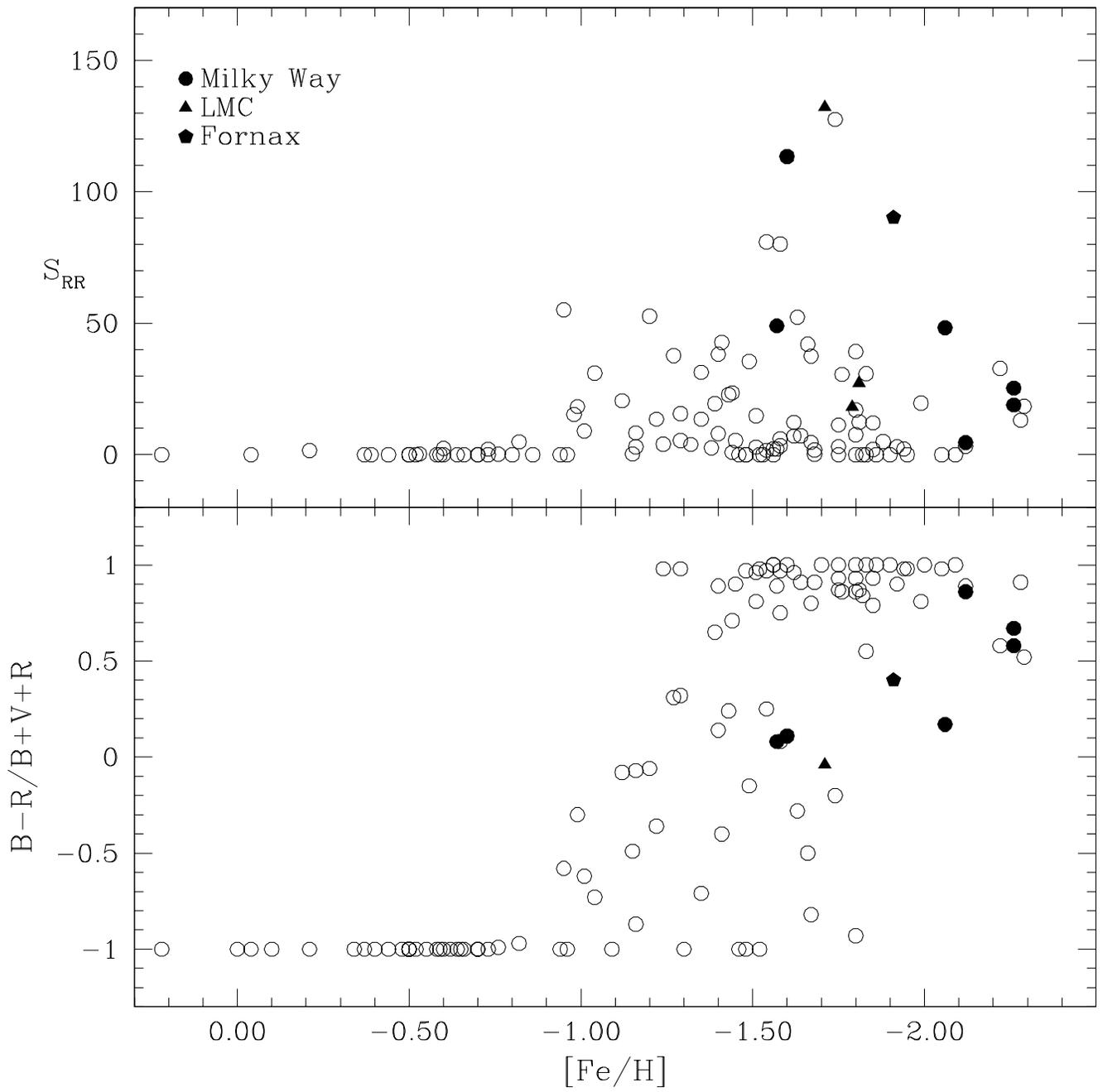
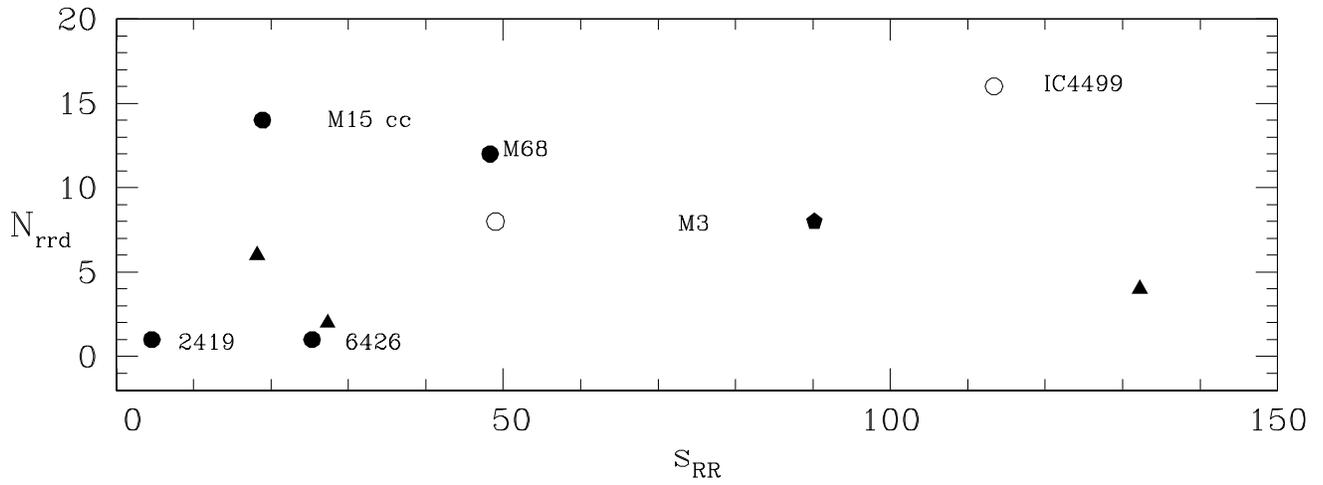
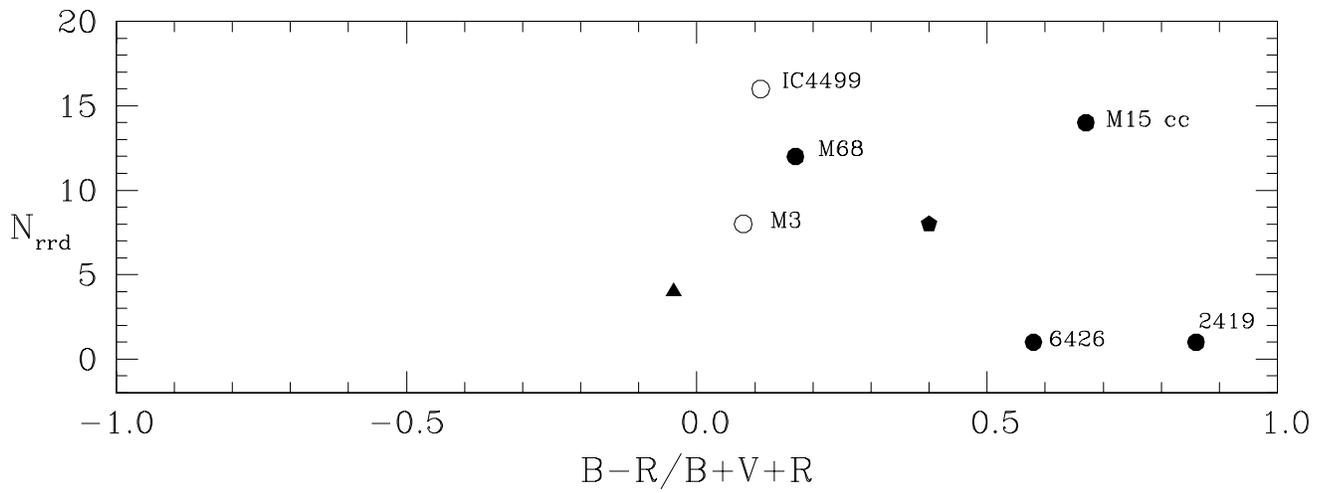
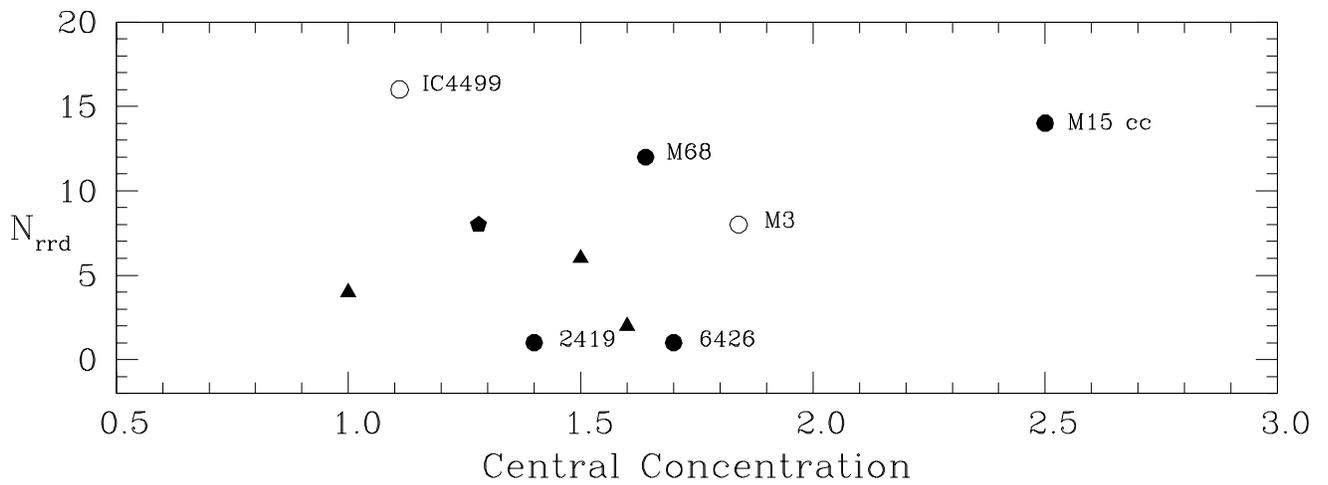
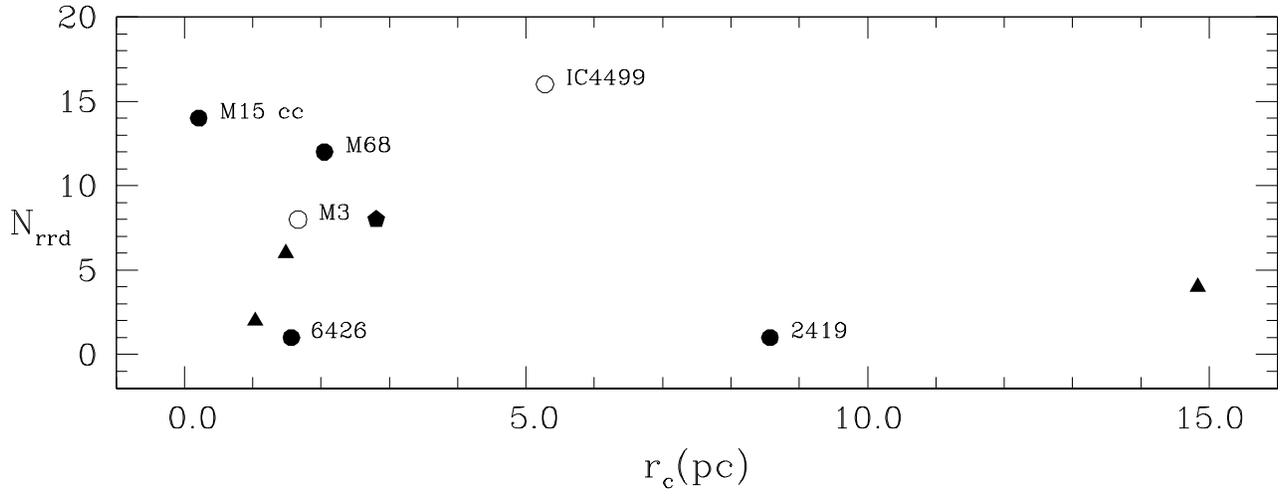
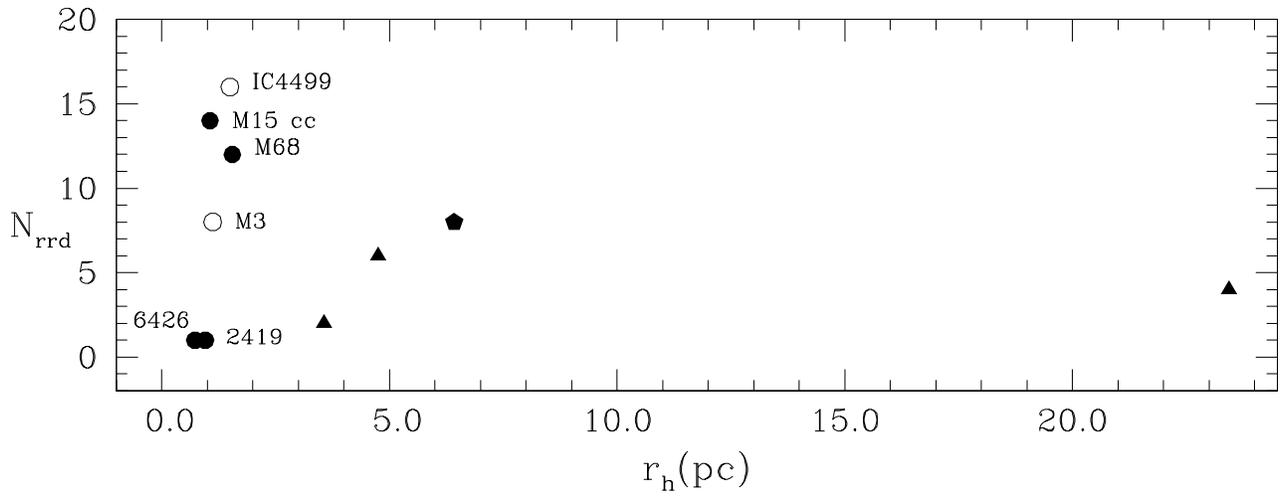
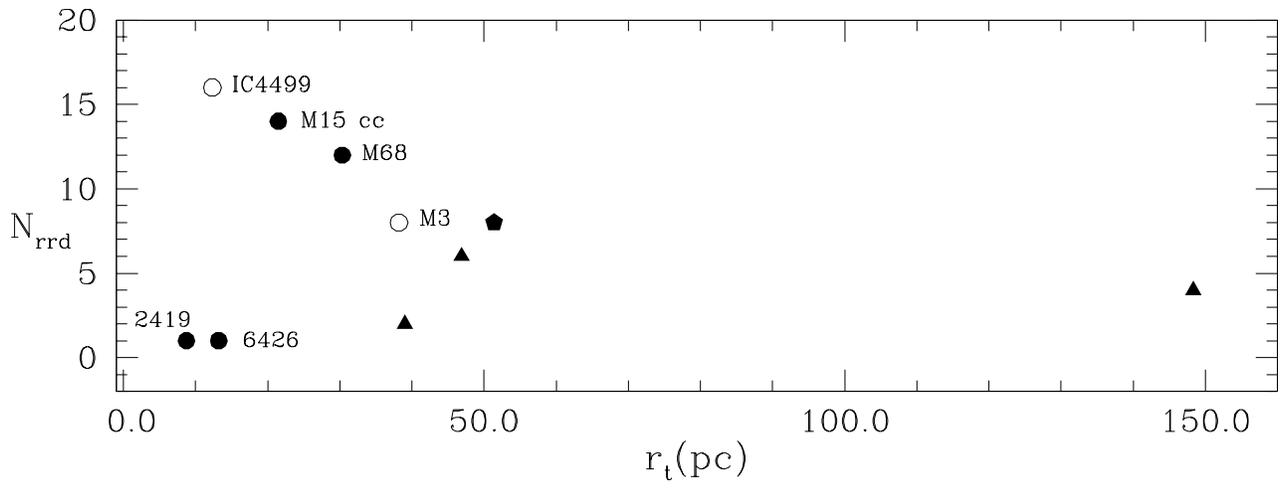
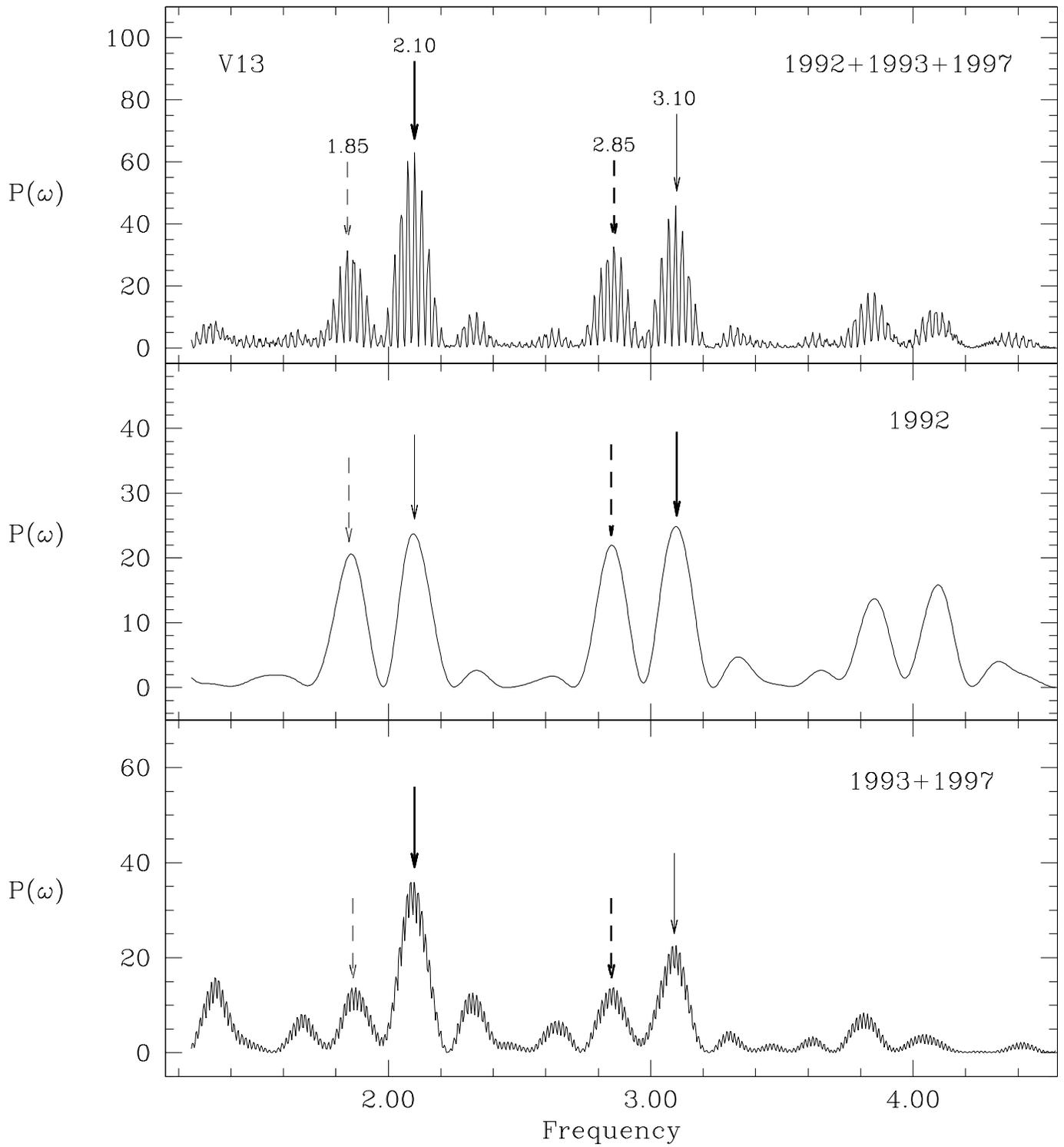
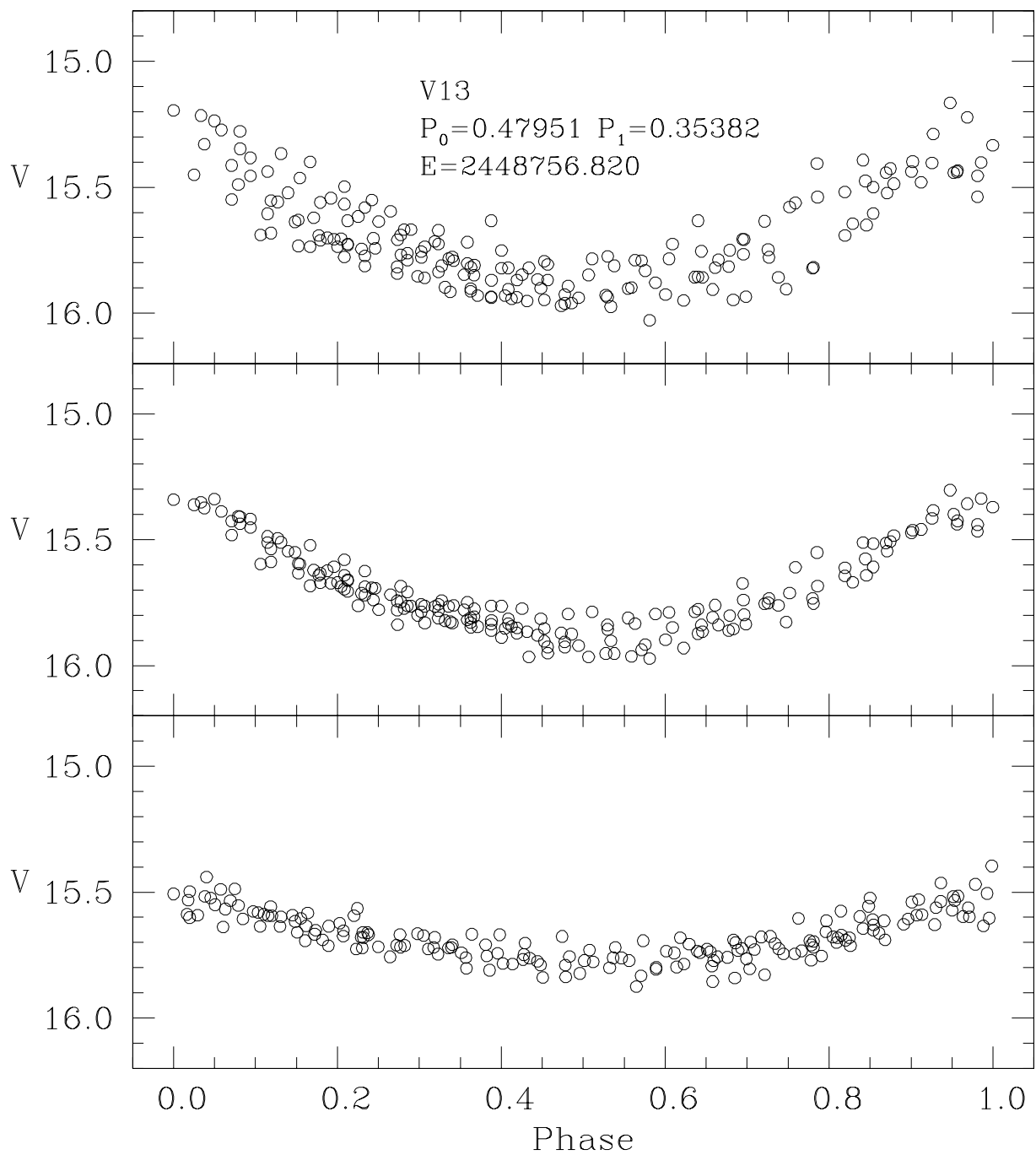
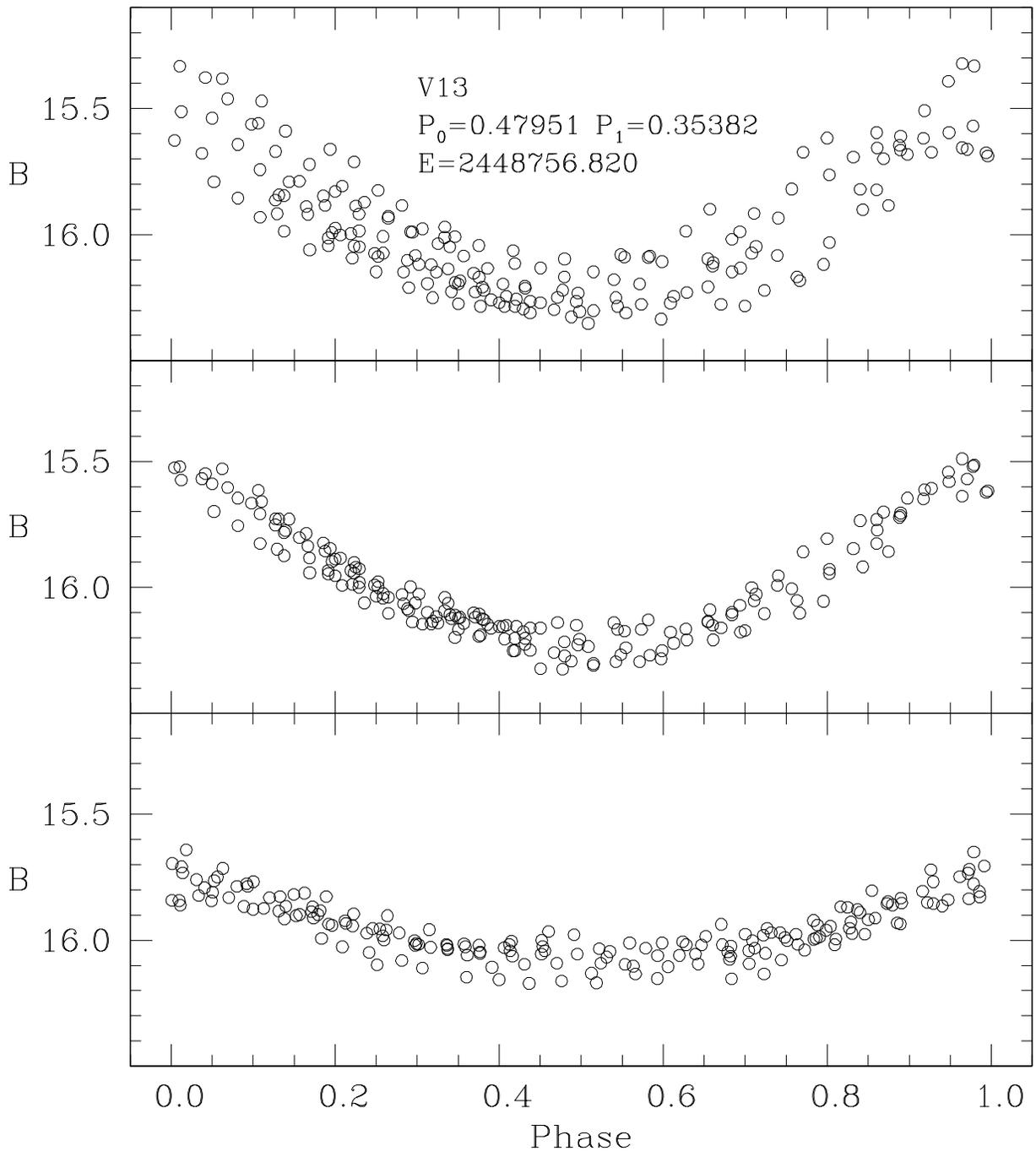
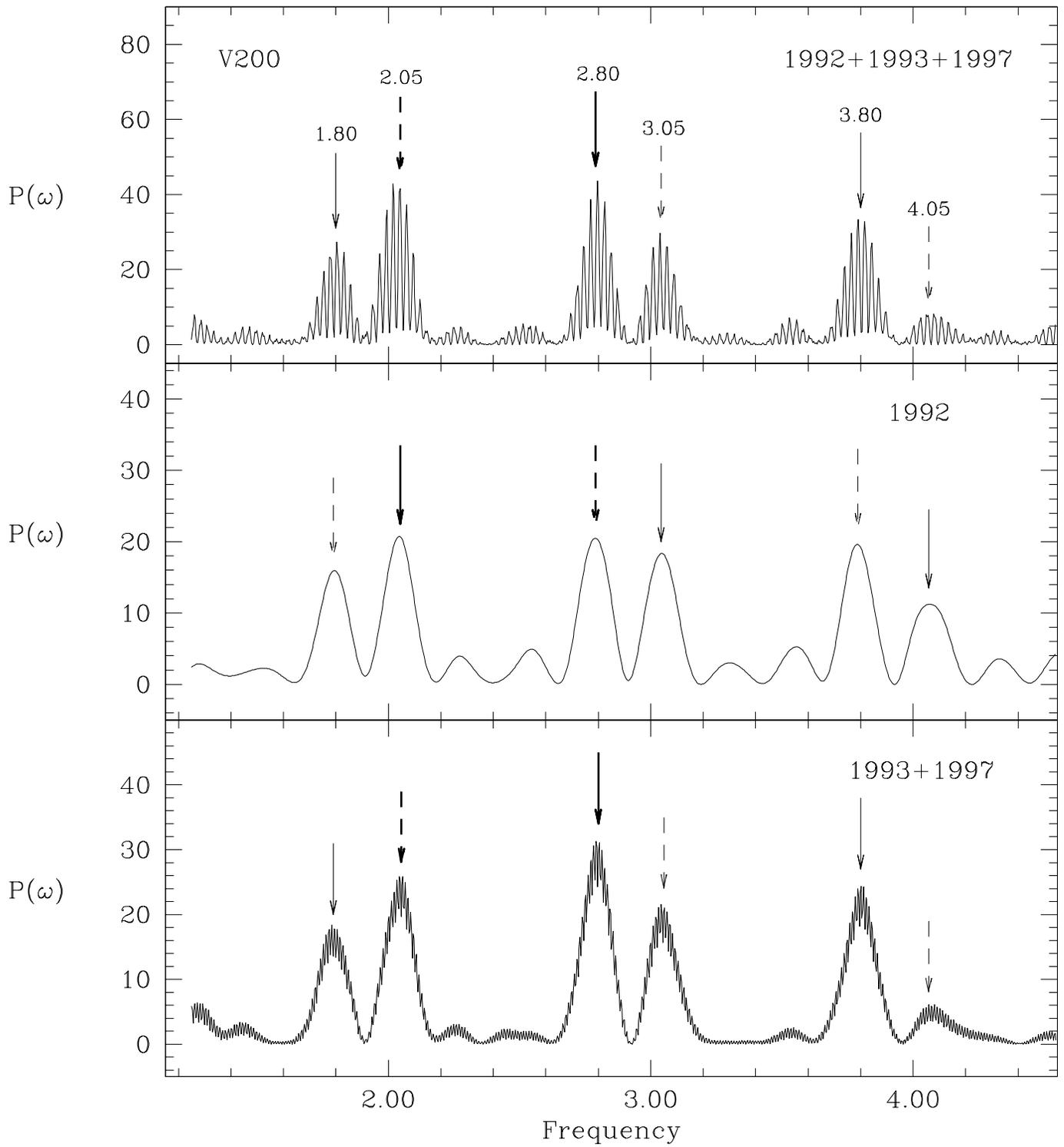
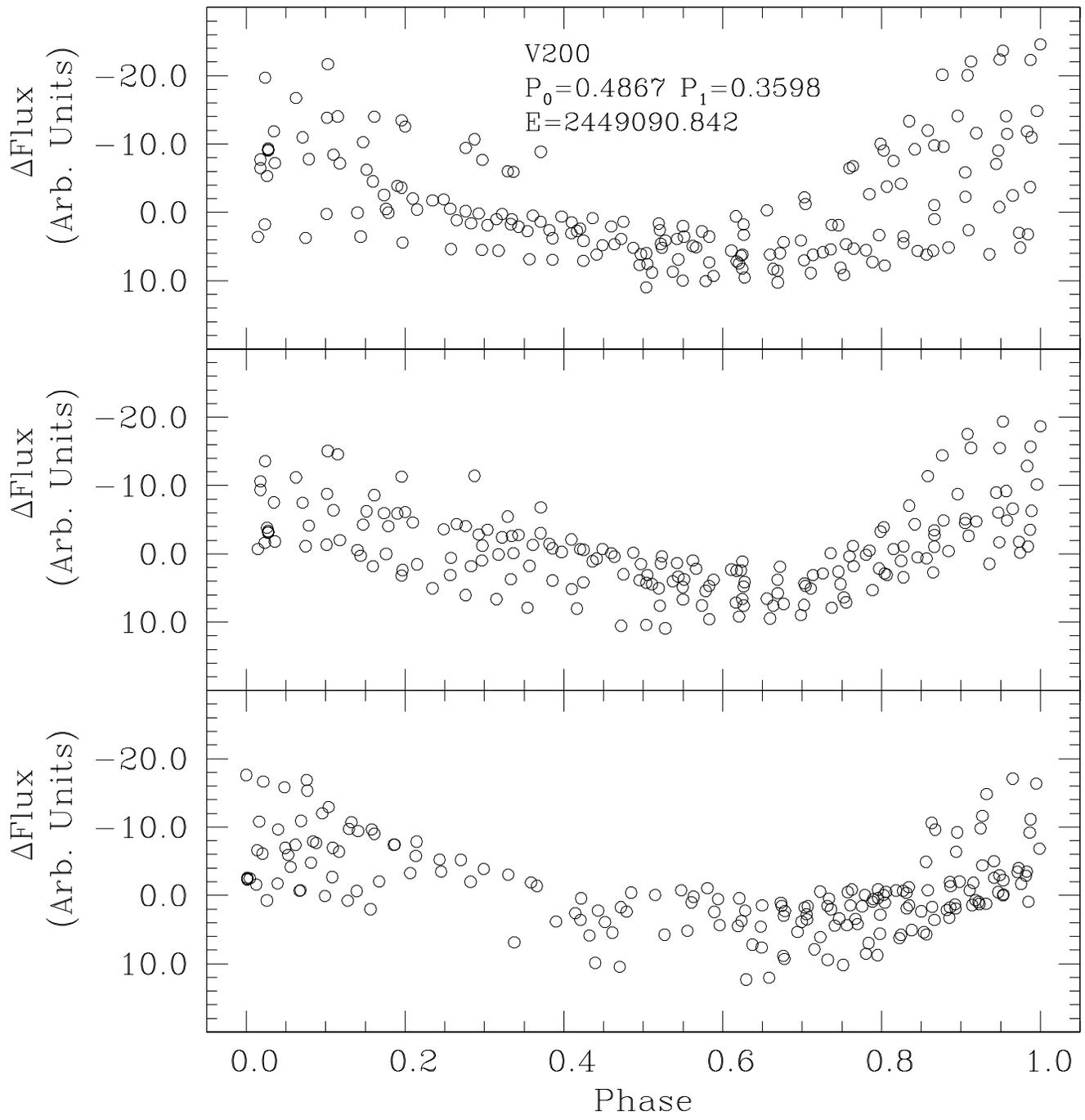
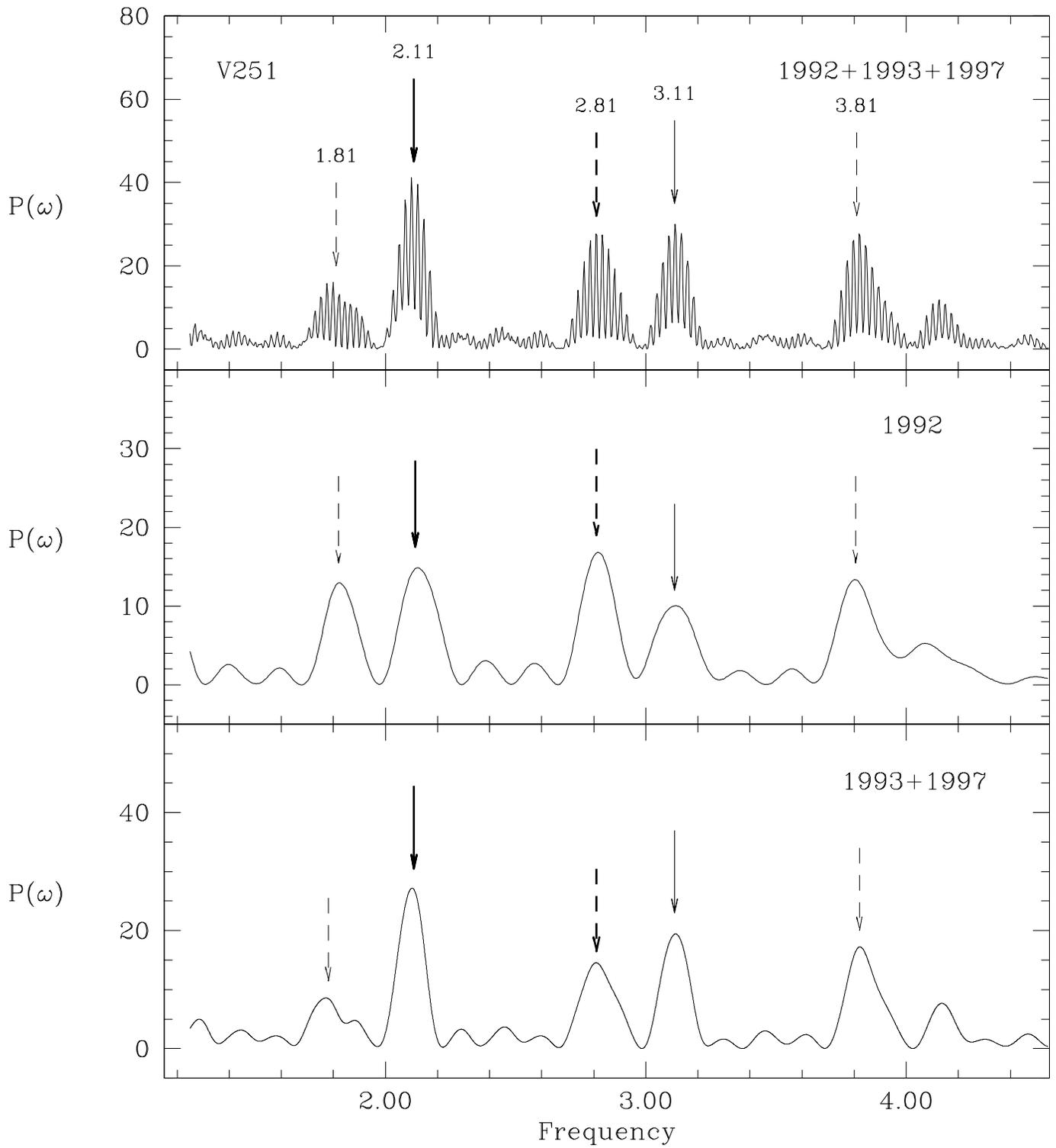
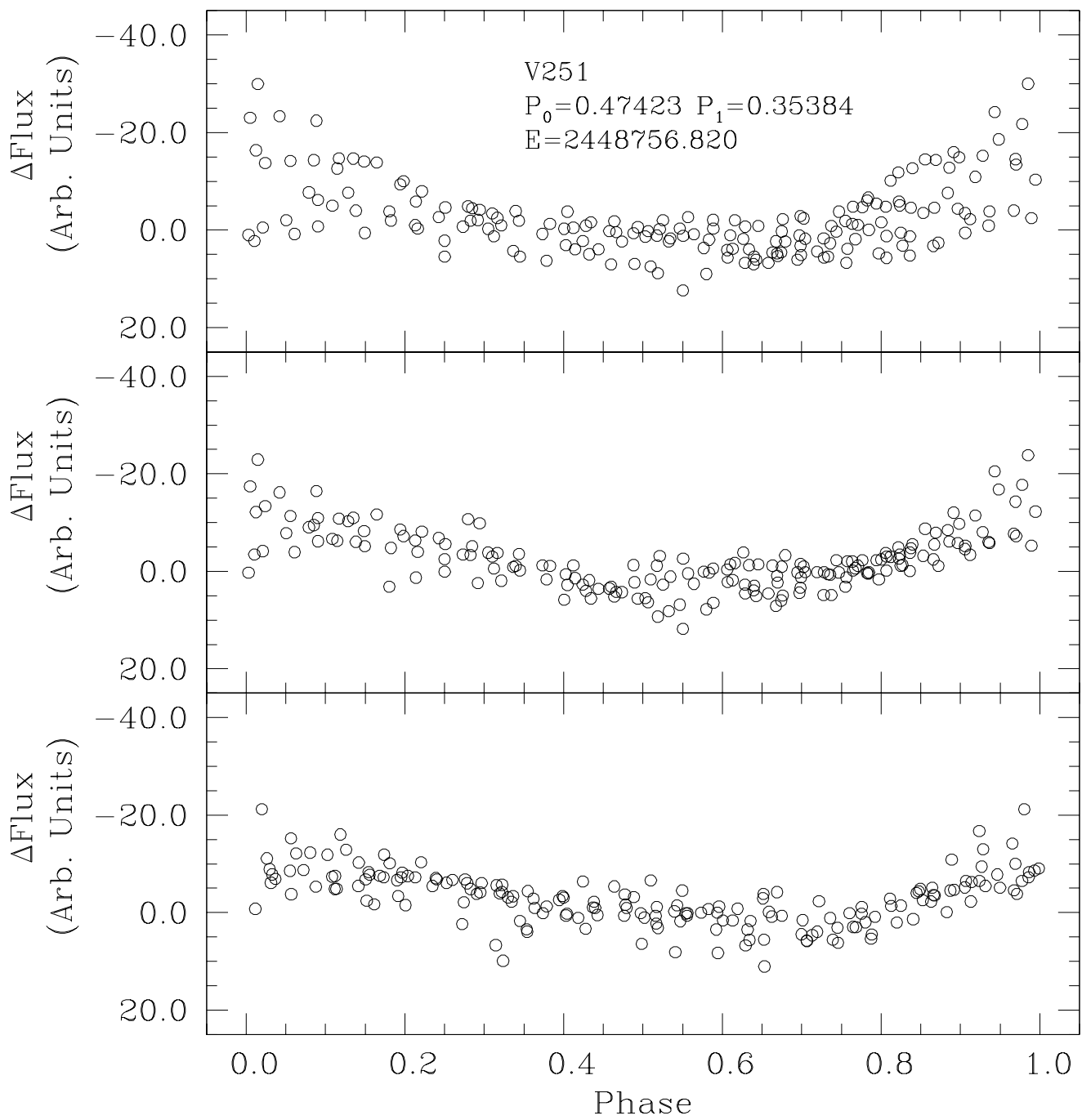
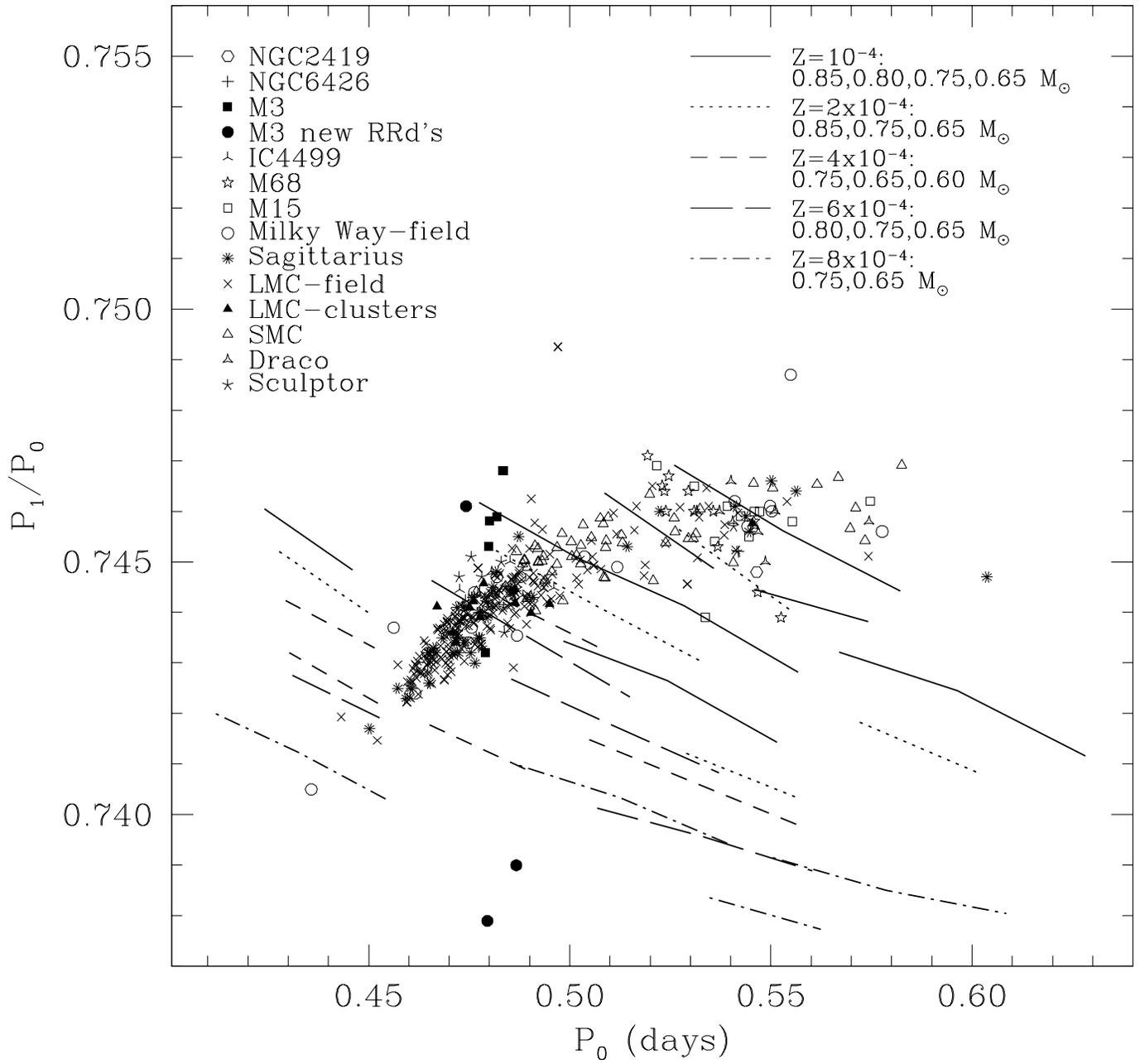
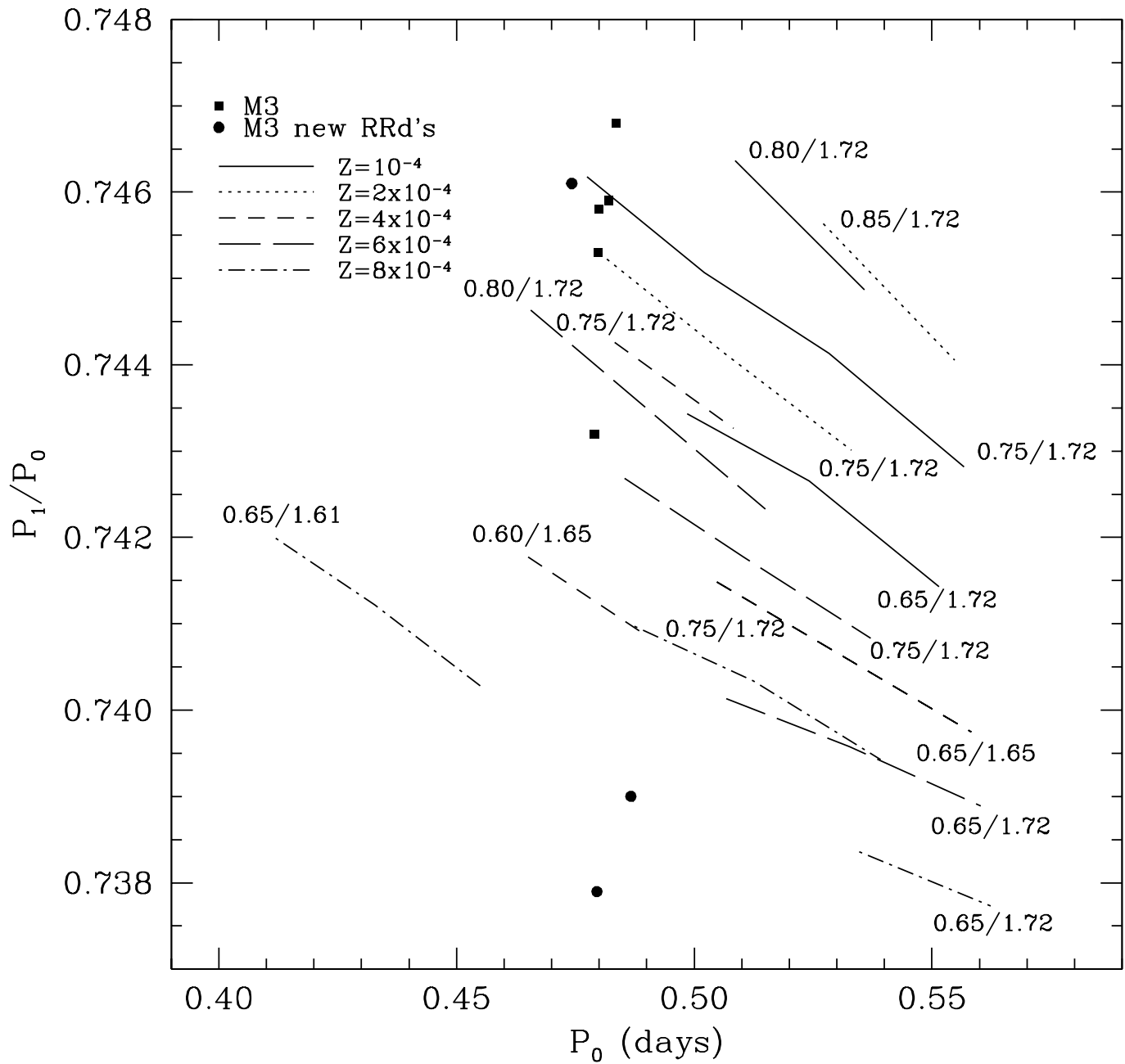
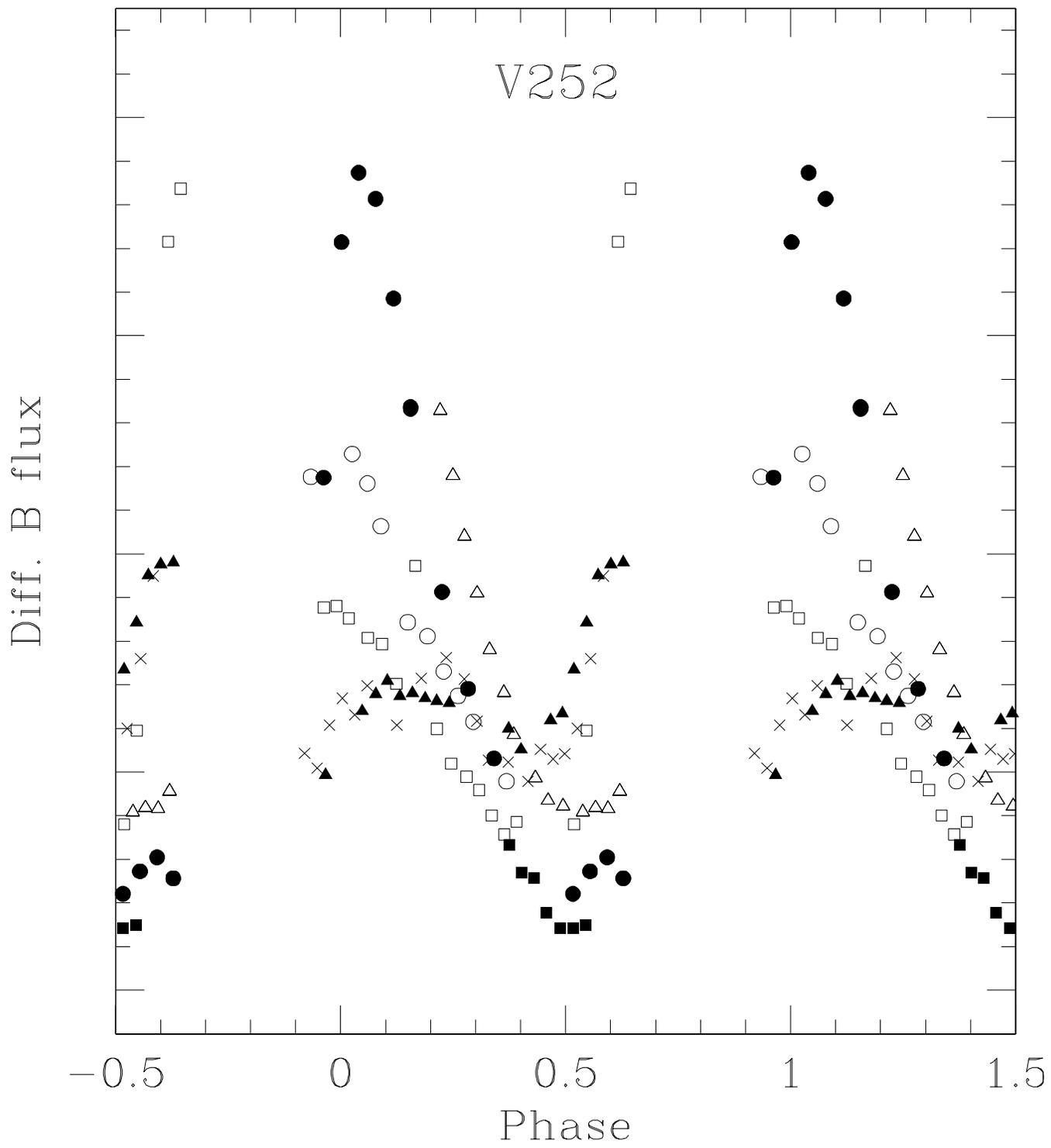
\end{document}